\documentclass[reprint,superscriptaddress,amsmath,amssymb,aps,pra]{revtex4-2}

\usepackage{balance}

\usepackage{color}
\usepackage{booktabs}
\usepackage{makecell}
\usepackage{tabularx}
\usepackage{array}
\usepackage{graphicx}
\usepackage{subfigure}
\usepackage{stfloats}
\usepackage{caption}
\usepackage{multirow}
\usepackage{mathtools}
\usepackage{amsthm}
\usepackage{dcolumn}
\usepackage{bm}
\usepackage{algorithm}
\usepackage{algpseudocode}
\usepackage[colorlinks,linkcolor=blue,anchorcolor=blue,citecolor=blue]{hyperref}
\usepackage[T1]{fontenc}

\usepackage{braket}

\begin{document}

\title{Characterizing Quantum Codes via the Coefficients in Knill-Laflamme Conditions}

\author{Mengxin Du\,}
\affiliation{Department of Physics, The University of Texas at Dallas, Richardson, Texas 75080, USA}

\author{Chao Zhang\,}
\affiliation{Department of Physics, The Hong Kong University of Science and Technology, Clear Water Bay, Kowloon, Hong Kong}

\author{Yiu Tung Poon\,}
\affiliation{Department of Mathematics, Iowa State University, Ames, Iowa 50011, USA}

\author{Bei Zeng\,}
\email{bei.zeng@utdallas.edu}
\affiliation{Department of Physics, The University of Texas at Dallas, Richardson, Texas 75080, USA}

\date{\today}

\begin{abstract}
Quantum error correction (QEC) is essential for protecting quantum information against noise, yet understanding the structure of the Knill-Laflamme (KL) coefficients \( \lambda_{ij} \) from the condition \( PE_i^\dagger E_j P = \lambda_{ij} P \) remains challenging, particularly for nonadditive codes. In this work, we introduce the signature vector \( \vec{\lambda}(P) \), composed of the off-diagonal KL coefficients \( \lambda_{ij} \), where each coefficient corresponds to equivalence classes of errors counted only once. We define its Euclidean norm \( \lambda^*(P) \) as a scalar measure representing the total strength of error correlations within the code subspace defined by the projector \( P \). We parameterize \( P \) on a Stiefel manifold and formulate an optimization problem based on the KL conditions to systematically explore possible values of \( \lambda^* \). Moreover, we show that, for \( ((n,K,d)) \) codes, \( \lambda^* \) is invariant under local unitary transformations. Applying our approach to the \(((6, 2, 3))\) quantum code, we find that \( \lambda^*_{\text{min}} = \sqrt{0.6} \) and \( \lambda^*_{\text{max}} = 1 \), with \( \lambda^* = 1 \) corresponding to a known degenerate stabilizer code. We construct continuous families of new nonadditive codes parameterized by vectors in \( \mathbb{R}^5 \), with \( \lambda^* \) varying over the interval \( [\sqrt{0.6}, 1] \). For the \(((7, 2, 3))\) code, we identify \( \lambda^*_{\text{min}} = 0 \) (corresponding to the non-degenerate Steane code) and \( \lambda^*_{\text{max}} = \sqrt{7} \) (corresponding to the permutation-invariant code by Pollatsek and Ruskai), and we demonstrate continuous paths connecting these extremes via cyclic codes characterized solely by \( \lambda^* \). Our findings provide new insights into the structure of quantum codes, advance the theoretical foundations of QEC, and open new avenues for investigating intricate relationships between code subspaces and error correlations.
\end{abstract}

\maketitle

\section{Introduction}

Quantum error correction (QEC) is essential for protecting quantum information from the noise and errors that inevitably arise in quantum systems~\cite{calderbank1996good,nielsen2010quantum,gottesman1997stabilizer,steane1996error,lidar2013quantum,kitaev1997quantum}. A deeper understanding of the structure of the set given by all possible coefficients \( \lambda_{ij} \), which arise from the Knill-Laflamme (KL) conditions \( PE_i^\dagger E_jP = \lambda_{ij}P \)~\cite{knill1997theory}, can provide valuable insights into the performance and underlying properties of quantum error-correcting codes. However, achieving this understanding is challenging. Nonadditive codes, which lie outside the stabilizer formalism, are particularly difficult to analyze, as relatively few examples have been systematically studied~\cite{calderbank1998quantum,rains1997nonadditive,cross2008codeword}. Moreover, degenerate codes—where multiple errors produce the same effect on the code space—exhibit inherently quantum phenomena, such as overlapping error syndromes, that lack classical analogues and remain poorly understood~\cite{lidar2003decoherence,sarvepalli2010degenerate}. These complexities make it difficult to systematically explore the structure of the set of all possible $\lambda_{ij}$ values. As a result, there is currently no comprehensive framework for understanding the distribution of these coefficients, leaving important questions about their structure and implications for quantum error correction unanswered.

We analyze the structure defined by the set of all possible values of \( \lambda_{ij} \) that satisfy the KL conditions, which govern how pairs of errors interact within the code subspace defined by \( P \). To encapsulate these interactions, we introduce signature vector \( \vec{\lambda}(P) \), composed of the off-diagonal elements \( \lambda_{ij} \) (with each coefficient corresponding to equivalent errors counted only once), which capture the non-trivial correlations between errors. The overall strength of these interactions is quantified by \( \lambda^*(P) \), the Euclidean norm of the signature vector. This scalar value provides a measure of the total strength of error interactions within the code subspace, offering a new perspective on the role of these interactions in QEC. Crucially, for \( ((n,K,d)) \) codes, \( \lambda^* \) is a function of the purity of the local reduced density matrices (RDMs) of the codewords, making it invariant under local unitary operations. This local unitary invariance allows \( \lambda^* \) to serve as a powerful tool for distinguishing locally unitary inequivalent quantum codes and identifying different codes based on their error interaction structures.

The focus of this paper is to study the range of \( \lambda^* \): to understand the minimum and maximum values of \( \lambda^* \) (denoted by \( \lambda^*_{\text{min}} \) and \( \lambda^*_{\text{max}} \)), and to determine whether the range of \( \lambda^* \) is continuous between these extrema. We propose an algorithm to find \( \lambda^*_{\text{min}} \) and \( \lambda^*_{\text{max}} \), as well as to identify possible intermediate values of \( \lambda^* \) between these limits. The core of our method involves characterizing the projector \( P \) on a Stiefel manifold, which provides a natural parameterization of the code subspace. We then formulate an optimization problem by constructing a loss function based on the KL conditions. This approach allows us to systematically explore the set of all possible \( \lambda_{ij} \) values and identify various values of \( \lambda^* \) corresponding to different quantum codes.

Using our method, we find that for the \(((6, 2, 3))\) quantum code, \( \lambda^*_{\text{min}} = \sqrt{0.6} \) and \( \lambda^*_{\text{max}} = 1 \). The value \( \lambda^* = 1 \) corresponds to the degenerate stabilizer code described in~\cite{lidar2003decoherence}. However, there are no known codes corresponding to \( \lambda^* =\sqrt{0.6} \). We construct families of nonadditive codes, parameterized by four mutually orthogonal real vectors \( a, b, c, d \in \mathbb{R}^5 \), with \( \lambda^* \) parameterized by a vector \( e \in \mathbb{R}^5 \), orthogonal to \( a, b, c, d \), which varies continuously over the interval \( [\sqrt{0.6}, 1] \). This confirms that the range of \( \lambda^* \) for the \(((6, 2, 3))\) code is indeed \( [\sqrt{0.6}, 1] \). For each distinct value of \( \lambda^* \in [\sqrt{0.6}, 1] \), our construction yields locally inequivalent codes, parameterized by the vector \( e \in \mathbb{R}^5 \).

For the \(((7, 2, 3))\) code, we find that $\lambda^*_{\text{min}}=0$ and
\( \lambda^*_{\text{max}}=\sqrt{7} \), where \( \lambda^* = 0 \) corresponds to the non-degenerate Steane code~\cite{steane1996simple,steane1996multiple}, and \( \lambda^* = \sqrt{7} \) corresponds to the permutation-invariant code proposed by Pollatsek and Ruskai~\cite{pollatsek2004permutationally}.
We identify families of cyclic \(((7, 2, 3))\) codes that trace continuous paths in the solution space. These paths, each characterized by a single parameter, which is simply \( \lambda^* \), with $\lambda^*$ varies continuously over the interval $[0,\sqrt{7}]$, directly connecting the Steane code and the permutation-invariant code. This finding demonstrates that it is possible to smoothly connect these two distinct codes while preserving cyclic symmetry, offering new insights into the relationship between different locally inequivalent quantum codes and their symmetry properties.

Our approach offers a systematic method to explore the range of \( \lambda^* \), resulting in the construction of numerous new nonadditive codes for \( ((6,2,3)) \) and \( ((7,2,3)) \), with \( \lambda^* \) varying continuously from \( \lambda^*_{\text{min}} \) to \( \lambda^*_{\text{max}} \). The ability to identify and quantify the range of \( \lambda^* \) provides novel insights into the structure of quantum codes, particularly in nonadditive cases. This framework opens new avenues for investigating the intricate relationships between code subspaces and error interactions, offering a deeper understanding of the mathematical structure underlying quantum error correction.

We organize our paper as follows. In Section~\ref{pre}, we discuss preliminaries on quantum error correction and code parameters. In Section~\ref{sec:structure}, we define the signature vector and its norm \( \lambda^* \), show that \( \lambda^* \) is invariant under local unitary operations by linking it to the purity of the RDMs of codewords, and develop an algorithm to find the maximum and minimum values of \( \lambda^* \). In Sections~\ref{sec:6-2-3} and~\ref{sec:7-2-3}, we apply our method to the \(((6, 2, 3))\) and \(((7, 2, 3))\) quantum codes, respectively, demonstrating how \( \lambda^* \) varies and constructing new nonadditive codes.

\section{Preliminary}\label{pre}

In quantum error correction, the goal is to protect quantum information from errors caused by a noisy quantum channel. Quantum error-correcting codes (QECCs) are constructed to correct a specified set of errors. The Knill-Laflamme (KL) condition for quantum error correction can be expressed as:

\begin{equation}
P E_i^\dagger E_j P = \lambda_{ij} P, \quad \forall i,j,
\label{eq:knill-laflamme}
\end{equation}

\noindent where \( P \) denotes the projector onto the code subspace, \( E_i \) and \( E_j \) represent the Kraus operators corresponding to the possible errors, and \( \lambda_{ij} \) are complex scalars that characterize how the pair of errors \( E_i \) and \( E_j \) interact within the code subspace. This condition ensures that errors are correctable, provided that they act within the designated subspace and satisfy this equation.

The dimension of the code subspace is denoted as \( K \), and if the logical information is encoded in a subspace of \( K \)-dimensional logical qubits within an \( n \)-dimensional physical qubit system, then \( P \) is an \( n \times n \) matrix, and its rank equals \( K \). The code subspace \( \mathcal{C} \) can be written as the span of orthonormal basis vectors \( \{ |\psi_1\rangle, |\psi_2\rangle, \dots, |\psi_K\rangle \} \), which span the logical space. The projector onto the code subspace is given by

\begin{equation}
P = \sum_{i=1}^{K} |\psi_i\rangle \langle \psi_i|.
\label{eq:projector}
\end{equation}
\noindent The quantum error correction condition can then be expressed in terms of the basis vectors spanning the code subspace:

\begin{equation}
\langle \psi_k | E_i^\dagger E_j | \psi_l \rangle = \lambda_{ij} \delta_{kl}, \quad \forall i,j
\label{eq:qec-condition}
\end{equation}
where the scalars \( \lambda_{ij} \) describe how the errors \( E_i \) and \( E_j \) affect the code subspace.

A \textit{non-degenerate QECC} is characterized by the Hermitian matrix \( \lambda_{ij} \) being non-singular (having full rank), which means that the determinant of \( \lambda_{ij} \) is non-zero and the matrix is invertible \cite{calderbank1996good}. This implies that all errors have distinct effects on the code space and can be uniquely identified and corrected. In contrast, a \textit{degenerate QECC} arises when the matrix \( \lambda_{ij} \) is singular (not of full rank), indicating that there are linear dependencies among the error operators when restricted to the code space \cite{steane1996error}. Some errors or combinations of errors may have the same effect on the code space, making them indistinguishable. A \textit{completely degenerate code}, or \textit{decoherence-free subspace (DFS)}, represents an extreme case where all \( \lambda_{ij} \) elements are equal, resulting in a matrix of rank $1$, meaning \( \lambda_{ij} = \lambda \) for all \( i, j \). In this scenario, the code space remains invariant under certain noise processes~\cite{lidar1998decoherence,lidar2003decoherence}.

An \(((n, K, d))\) quantum error-correcting code is defined by three key parameters: \( n \), the number of physical qubits used to encode the quantum information; \( K \), the dimension of the code space, which corresponds to the number of logical qubits the code can protect (for example, if \( K = 2^k \), the code protects \( k \) logical qubits); and \( d \), the distance of the code, which determines the minimum number of physical qubit errors required to cause a logical error. The distance \( d \) indicates the code’s ability to detect and correct errors. Specifically, an \(((n, K, d))\) code can detect up to \( d-1 \) qubit errors and correct up to \( t = \left\lfloor \frac{d-1}{2} \right\rfloor \) qubit errors \cite{calderbank1996good,steane1996error}. A well-known example is the Steane code, which is an \(((7, 2, 3))\) code. This code encodes one logical qubit into seven physical qubits and can correct up to one qubit error and detect up to two qubit errors \cite{steane1996error}.

Furthermore, quantum error-correcting codes may either be \textit{non-additive} or \textit{additive}. Non-additive codes are a generalization of stabilizer (additive) codes and allow encoding of quantum information without adhering to the \( 2^k \) constraint for the dimension \( K \). The code distance \( d \), which is the minimum weight of an undetectable error, remains critical in both types of codes, as it determines how many errors can be detected and corrected.

Two quantum error-correcting codes \( P_1 \) and \( P_2 \) are locally equivalent if one can be transformed into the other by local unitary operations or local Clifford operations applied to individual qubits. Formally, \( P_1 \) and \( P_2 \) are locally equivalent if there exists a unitary transformation \( U = U_1 \otimes U_2 \otimes \dots \otimes U_n \), where each \( U_i \) acts on a single qubit, such that \( P_2 = U P_1 U^\dagger \). This local equivalence ensures that the overall structure of the code and parameters \( n \), \( K \), \( d \) are preserved, even though individual states within the code space may change under the transformation \cite{gottesman1997stabilizer}.

In practice, to test whether two \(((n, K, d))\) codes \( P_1 \) and \( P_2 \) are local unitary equivalent, we can use \textit{Quantum weight enumerators}~\cite{shor1997quantum}, which were defined by
	\begin{equation}
		A(z) = \sum_{j=0}^{n} A_j z^j, \quad B(z) = \sum_{j=0}^{n} B_j z^j
	\end{equation}
	with coefficients
	\begin{equation}\label{eq:qwe-A}
		A_j = \frac{1}{K^2}\sum_{\operatorname{wt}(O_{\alpha})=j} \operatorname{Tr}(O_{\alpha}P_c)\operatorname{Tr}(O_{\alpha}^{\dagger}P_c),
	\end{equation}
	\begin{equation}
		B_j = \frac{1}{K}\sum_{\operatorname{wt}(O_{\alpha})=j} \operatorname{Tr}(O_{\alpha}P_cO_{\alpha}^{\dagger}P_c).\mkern44mu
	\end{equation}

Here
	\begin{equation}
		O_{\alpha} \in \{X,Y,Z,I\}^{\otimes n}
	\end{equation}
 are $n$-fold Pauli tensor product.
	Denote the number of $X$ factors, $Y$ factors and $Z$ factors in $O_{\alpha}$ as $\operatorname{wt}_{\mathrm{X}}(O_{\alpha})$, $\operatorname{wt}_{\mathrm{Y}}(O_{\alpha})$, and $\operatorname{wt}_{\mathrm{Z}}(O_{\alpha})$. The weight of $O_{\alpha}$ is
	\begin{equation}
		\operatorname{wt}(O_{\alpha}) = \operatorname{wt}_{\mathrm{X}}(O_{\alpha}) + \operatorname{wt}_{\mathrm{Y}}(O_{\alpha}) + \operatorname{wt}_{\mathrm{Z}}(O_{\alpha}).
	\end{equation}
And other related concepts including Rains’ unitary and shadow quantum weight enumerators~\cite{rains1998quantum, rains1999quantum,
miller2024experimental}.

\section{Structure of $\lambda_{ij}$}\label{sec:structure}
Throughout this paper, we assume that the error operators \( E_i \) are Pauli operators for convenience. However, our method naturally extends to non-Pauli errors as well. In this context, we simplify the analysis by focusing on the case where \( i<j \). The quantum error correction criterion (KL) condition is given by
$
P E_i^\dagger E_j P = \lambda_{ij} P, \quad \forall i,j
$,
where \( P \) is the projector onto the code subspace, and \( \lambda_{ij} \) encodes the interaction between errors \( E_i \) and \( E_j \) on the code subspace.

\subsection{The signature vector}

To capture the nature of these off-diagonal interactions, we define the signature vector as off-diagonal elements in matrix \( \lambda_{ij}\) in KL conditions. However, when dealing with Pauli errors, the product \( E_i^\dagger E_j \) can be proportional to another Pauli operator that may already be included in our set of errors. This leads to a double counting problem, as the same operator can appear multiple times due to different pairs \( (i, j) \). For example, with single-qubit errors such as \( X_i \) and \( Y_i \), the product \( X_i^\dagger Y_i \) is proportional to \( Z_i \), which might already be included in the error set.

To resolve this issue and avoid double counting, we refine our definition by considering only unique error interactions. Specifically, we define an equivalence relation on the set of operator products \( E_i^\dagger E_j \), where two operators are considered equivalent if they are proportional up to a scalar multiple (including global phase). That is,
\[
E_i^\dagger E_j \sim E_k^\dagger E_l \quad \text{if} \quad E_i^\dagger E_j = \alpha E_k^\dagger E_l,
\]
for some non-zero \( \alpha \).

We then construct a set of representatives from each equivalence class of these operator products, ensuring that each unique operator (up to proportionality) is included only once. The \emph{Signature Vector} \( \vec{\lambda}(P) \) associated with a projector \( P \) is then defined using the corresponding \( \lambda_{ij} \) values for these representatives:
\begin{equation}
\vec{\lambda}(P) = (\lambda_{i_1 j_1}, \lambda_{i_2 j_2}, \dots, \lambda_{i_m j_m}),
\end{equation}
where each \( E_{i_k}^\dagger E_{j_k} \) is a distinct representative in \( \mathcal{S} \), and \( i_k < j_k \) for all \( k \).

Each component of \( \vec{\lambda}(P) \) represents the interaction between a pair of distinct errors \( E_{i_k} \) and \( E_{j_k} \) as captured by the quantum error correction criterion, without redundant counting. By focusing on these unique off-diagonal elements, the signature vector reflects the degree of correlation between different errors on the code space defined by \( P \) while avoiding double counting of equivalent error interactions.

Let \( \mathcal{W}_{\text{error}} \) denote the set of all possible signature vectors corresponding to different projectors \( P \) that satisfy the quantum error correction criterion for a given error model \( \{ E_i \} \). Formally, we define:
\begin{align}
\mathcal{W}_{\text{error}} = \big\{ \vec{\lambda}(P) : P \text{ satisfies the QEC} \notag \\
\text{criterion for error model } \{ E_i \}\big\}
\end{align}
The set \( \mathcal{W}_{\text{error}} \) represents all possible interactions between distinct errors under the given error model. Analyzing the structure of \( \mathcal{W}_{\text{error}} \) is crucial for understanding the properties of quantum error-correcting codes.

The structure of \( \mathcal{W}_{\text{error}} \) is closely related to the \textit{rank-\( K \) joint numerical range} of the set of operators \( \{ E_i^\dagger E_j \} \). The rank-\( K \) joint numerical range
for a set of operators $\{A_i\}_{i=1}^m$ is defined as:
\begin{align}
& W^{(K)}(A_1,A_2,\ldots, A_m) = \notag\\\Big\{ (\lambda_{i}) :
& \text{There exists a rank-} K \text{ projector } P \notag \\
& \text{such that } P A_i P = \lambda_{i} P \Big\}
\end{align}
This definition aligns with the quantum error correction conditions, therefore, studying \( \mathcal{W}_{\text{error}} \) is equivalent to analyzing the rank-\( K \) joint numerical range of the operators \( \{ E_i^\dagger E_j \} \).

In particular, for \( ((n,K,d)) \) codes, we have
\begin{eqnarray*}
\mathcal{W}_{\text{error}}&=&W^{(K)}(\{ E_{i_k}^\dagger E_{j_k} \}, i_k < j_k)\\
&=&W^{(K)}(\{O_{\alpha}\}, 0<\operatorname{wt}(O_{\alpha})<d).
\end{eqnarray*}
Here
	$O_{\alpha} \in \{X,Y,Z,I\}^{\otimes n}$
are $n$-fold Pauli tensor product.

The rank-1 joint numerical range is known to be always connected. It is convex for $m=2$ Hermitian operators  (as shown by the Toeplitz-Hausdorff theorem~\cite{toeplitz1918algebraische,hausdorff1919wertvorrat}), so does the case for rank-$K$~\cite{woerdeman2008higher,li2008convex}.
But for $m>2$ the properties of higher-rank joint numerical ranges are less well understood~\cite{choi2006higher,choi2008geometry,li2008higher,li2011gnr,gau2011higher}. For \( K \geq 2 \), the rank-\( K \) joint numerical range of a set of operators is generally non-convex and can exhibit a complex structure, including disconnected components. As an example, consider a case of two qubits with
$A_{1} = X \otimes I$,
$A_{2} = X \otimes Z$,
$A_{3} = Y \otimes I$,
$A_{4} = Y \otimes Z$,
$A_{5} = Z \otimes I$. 
It can be shown that the rank-2 joint numerical range contains only two point
\[
W^{(2)}(A_1,A_2,A_3,A_4,A_5)=\{(0,0,0,0,1),(0,0,0,0,-1)\},
\]
hence $W^{(2)}$ is disconnected.

In general, the connectedness of the rank-\( K \) joint numerical range depends on the specific operators involved. In this work, we study the connectedness of \( \mathcal{W}_{\text{error}} \) for the case of the error-correcting code of interest.

\subsection{The length of the signature vector \( \lambda^*(P) \)}

We are particularly interested in the \textit{length of the Signature Vector}, denoted by \( \lambda^*(P) \), which is defined as:
\begin{equation}
\lambda^*(P) = \|\vec{\lambda}(P)\|_2 = \sqrt{\sum_{i_k<j_k} \lambda_{i_kj_k}^2}
\label{eq:lambda_norm}
\end{equation}
This length provides a measure of the overall strength of the error interactions on the code subspace for the given projector \( P \) \cite{miller2024experimental}. A natural lower bound is $\lambda^*\geq 0$, where $\lambda^*= 0$ may be achieved when there exists codes with all $\lambda_{ij}=0,\ i\neq j$ (e.g. nondegenerate stablizer code). And for a given quantum channel a natural upper bound may be achieved when there exists codes with all $\lambda_{ij}=1$ (i.e. the DFS).

Note that \( \mathcal{W}_{\text{error}} \) may not be connected in general, implying that the set of possible values of \( \lambda^*(P) \) may also not be continuous. In this study, we focus on the structure of the values of \( \lambda^*(P) \), specifically investigating the minimum and maximum values of \( \lambda^* \), denoted by \( \lambda^*_{\text{min}} \) and \( \lambda^*_{\text{max}} \), respectively. Furthermore, we aim to determine whether the range of \( \lambda^* \) is continuous between these extrema. Although the continuity of the \( \lambda^* \) range does not imply the connectedness of \( \mathcal{W}_{\text{error}} \), studying this range provides valuable insights into the structure of \( \mathcal{W}_{\text{error}} \).

For \( ((n,K,d)) \) codes,  \( \lambda^* \) is the length of the vectors in $W^{(K)}(\{O_{\alpha}\}, 0<\operatorname{wt}(O_{\alpha})<d)$. Furthermore, we show that  \( \lambda^* \), as defined above, is local unitary invariant (LUI), by linking \( \lambda^* \) to the purity of the reduced density matrices (RDMs) of the codewords (the purity of the reduced density matrix (RDM) is LUI).

Given a quantum state \( |\psi\rangle \), the RDM for the \( i \)-th subsystem is defined as:
\[
\rho^{(i)} = \mathrm{Tr}_{(i)^c}[|\psi\rangle\langle\psi|],
\]
where \( \mathrm{Tr}_{(i)^c} \) denotes the partial trace over all subsystems except the \( i \)-th one. The purity of this RDM is given by:
\begin{equation}
\mathcal{P}(\rho^{(i)}) = \mathrm{Tr}[(\rho^{(i)})^2]
\label{eq:purity}
\end{equation}
and since purity is invariant under local unitary transformations, the purity for 1-RDM, 2-RDM, ..., and up to \((d-1)\)-RDM is also LUI.

Next, consider the vector \( \lambda^{(i)} = (\mathrm{Tr}[\rho^{(i)}X_i], \mathrm{Tr}[\rho^{(i)}Y_i], \mathrm{Tr}[\rho^{(i)}Z_i]) \), which captures how the \( i \)-th subsystem interacts with the Pauli operators. The length of this vector is LUI, and is expressed as:
\[
\|\lambda^{(i)}\|_2 = \sqrt{\sum_{\alpha=1}^{3} (\lambda^{(i)}_\alpha)^2}.
\]
This can be rewritten in terms of the purity as \( \|\lambda^{(i)}\|_2^2 = 2 \mathrm{Tr}[\rho^{(i)}\rho^{(i)}] - 1 \), demonstrating that \( \|\lambda^{(i)}\|_2 \) is LUI.

Now, let \( \lambda^{(ij)} \) be a vector with 9 components, corresponding to the two-qubit interactions:
\[
\lambda^{(ij)} = \left( \mathrm{Tr}[\rho^{(ij)}X_iX_j], \mathrm{Tr}[\rho^{(ij)}X_iY_j], \cdots, \mathrm{Tr}[\rho^{(ij)}Z_iZ_j] \right).
\]
The length of this vector is also LUI, and is given by:
\[
\|\lambda^{(ij)}\|_2 = \sqrt{\sum_{\alpha=1}^{9} (\lambda^{(ij)}_\alpha)^2}.
\]
We can express this as:
\[
\|\lambda^{(ij)}\|_2^2 = 4 \mathrm{Tr}[\rho^{(ij)}\rho^{(ij)}] - 1 - \|\lambda^{(i)}\|_2^2 - \|\lambda^{(j)}\|_2^2,
\]
where each term on the right-hand side has already been shown to be LUI.

With invariance of weight-1 and weight-2 vectors, in a similar fashion, the length of weight-\((d-1)\) vectors \( \|\lambda^{(ij\cdots)}\|_2 \) can be proven to be LUI. Consequently, the length of the signature vector is given by:
\begin{align*}
(\lambda^*)^2 &= \sum_{i}\|\lambda^{(i)}\|_2^2 + \sum_{ij}\|\lambda^{(ij)}\|_2^2 \\
&\quad + \cdots + \text{(weight-$(d-1)$ term)},
\end{align*}
which is also LUI, as all terms involved are LUI. The LUI property can also be observed from the connection with quantum weight enumerators in Eq.(\ref{eq:qwe-A}). E.g., when $d=3$, $\lambda^{*2}=\sum_{i}\|\lambda^{(i)}\|_2^2+\sum_{ij}\|\lambda^{(ij)}\|_2^2=A_1+A_2$.

Since \( \lambda^* \) is local unitary invariant (LUI), it follows that if two quantum codes \( P_1 \) and \( P_2 \) correspond to different values of \( \lambda^* \), i.e., \( \lambda^*(P_1) \neq \lambda^*(P_2) \), then the two codes must be local unitary inequivalent. This means that the distinct values of \( \lambda^* \) reflect different structures in the code subspaces that cannot be transformed into one another via local unitary operations. This shows that \( \lambda^* \) serves as a useful tool for distinguishing some local unitary inequivalent codes. However, the converse does not hold: two local unitary inequivalent codes may correspond to the same value of \( \lambda^* \).

\subsection{Algorithm for calculating range of \( \lambda^* \)}
\label{method}

To parameterize the code space $P$, we use Stiefel manifold:
\[ \mathrm{St}\left(m,n\right)=\left\{ x\in\mathbb{C}^{m\times n}:m\geq n,x^{\dagger}x=I_{n}\right\}. \]
Parametrization for Stiefel manifold is given by:
\[ f(\theta)=\theta\left(\theta^{\dagger}\theta\right)^{-1/2}:\mathbb{C}^{m\times n}\to\mathrm{St}\left(m,n\right) \]
Above is the polar decomposition which maps (full rank) complex matrix $\theta\in\mathbb{C}^{m\times n}$ to a Stiefel matrix and all Stiefel matrices can be genrated in such a way \cite{zhu2024unifiedframeworkcalculatingconvex}. We embed the code subspace into Stiefel manifold:
\[ |\psi\rangle=\left\{ |\psi_{i}\rangle:i=1,\cdots,K\right\} \in\mathrm{St}\left(2^{n},K\right)\subseteq\mathbb{C}^{2^{n}\times K}. \]

For the parametrized states $|\psi\rangle$ (not a valid code yet), we can calculate the tensor $\tilde{\lambda}_{\alpha,i,j}=\langle\psi_{i}|O_{\alpha}|\psi_{j}\rangle$. For the subspace to be a valid code, the following loss term $\mathcal{L}_{\mathrm{KL}}$ should be optimized to zero
\[ \left\langle \tilde{\lambda}_{\alpha,i,i}\right\rangle _{i}=K^{-1}\sum_{i}\tilde{\lambda}_{\alpha,i,i},\;\left\Vert \tilde{\lambda}\right\Vert _{2}=\sqrt{\sum_{\alpha}\left\langle \tilde{\lambda}_{\alpha,i,i}\right\rangle _{i}^{2}} \]
\[ \mathcal{L}_{\mathrm{KL}}\left(\theta\right)=\sum_{\alpha,i\ne j}\left|\tilde{\lambda}_{\alpha,i,j}\right|^{2}+\sum_{\alpha,i}\left(\tilde{\lambda}_{\alpha,i,i}-\left\langle \tilde{\lambda}_{\alpha,i,i}\right\rangle _{i}\right)^{2} \]

To find the minimum length of $\lambda$ vector, we can optimize the following loss
\begin{equation}\label{eq:lambda-min}
    \mathcal{L}_{\underline{\ensuremath{\lambda}}}\left(\theta;\mu\right)=\mu\mathcal{L}_{\mathrm{KL}}+\left\Vert \tilde{\lambda}\right\Vert _{2}^{2}
\end{equation}
with $\mathcal{L}_{\mathrm{KL}}$ added as penalty and the hyper-parameter $\mu$ control the penalty strength. For a large enough $\mu$, the optimal value of $\mathcal{L}_{\underline{\lambda}}$ should corresponds to $\lambda$ with minimum length.

Similarly, to find the maximal length of $\lambda$, we can optimize the following loss function:
\begin{equation}
    \mathcal{L}_{\overline{\lambda}}\left(\theta;\mu\right)=\mu\mathcal{L}_{\mathrm{KL}}-\left\Vert \tilde{\lambda}\right\Vert _{2}^{2}.
\end{equation}

To find whether a code exists with length of $\lambda$ equal to $\lambda^*$, we can define such a loss function:
\begin{equation}\label{eq:lambda-length}
    \mathcal{L}\left(\theta;\mu,\lambda^{*}\right)=\mu\mathcal{L}_{\mathrm{KL}}+\left(\left\Vert \tilde{\lambda}\right\Vert_{2}^2-\lambda^{*2}\right)^{2}.
\end{equation}

Notice that similarly one can also find the code with a predefined vector $\vec{\lambda}$, just choose the loss function as:
\begin{equation}
    \mathcal{L}\left(\theta;\mu,\vec{\lambda}\right)=\mu\mathcal{L}_{\mathrm{KL}}+\left\Vert \tilde{\lambda}-\vec{\lambda}\right\Vert _{2}^{2}.
\end{equation}

\section{The $((6,2,3))$ codes}
\label{sec:6-2-3}

It is well known that $((5,2,3))$ code is unique up to local unitary equivalence,
with signature vector $\vec{\lambda}=0$, hence the range of $\lambda^*$ is a single point $0$. Much less is known about the range of $\lambda^*$ for the case of $((6,2,3))$. For stabilizer codes, there are only degenerate ones, for example the stabilizer code given in ~\cite{Shaw_2008}, with stabilizers given by
\[
\begin{array}{r@{\hspace{0.5em}}ccccccc}
g_1 & : & Y & I & Z & X & X & Y \\
g_2 & : & Z & X & I & I & X & Z \\
g_3 & : & I & Z & X & X & X & X \\
g_4 & : & I & I & I & Z & I & Z \\
g_5 & : & Z & Z & Z & I & Z & I \\
\end{array}
\]
For this code, all components of signature vector are zero except the term
$ \left\langle 0_{L}\right|Z_{4}Z_{6}\left|0_{L}\right\rangle =1$, hence
$\lambda^*=1$. All the other $((6,2,3))$ codes found in~\cite{cao2022quantum} also have $\lambda^*=1$.

\begin{figure}[h]
    \centering
    \includegraphics[width=0.9\columnwidth]{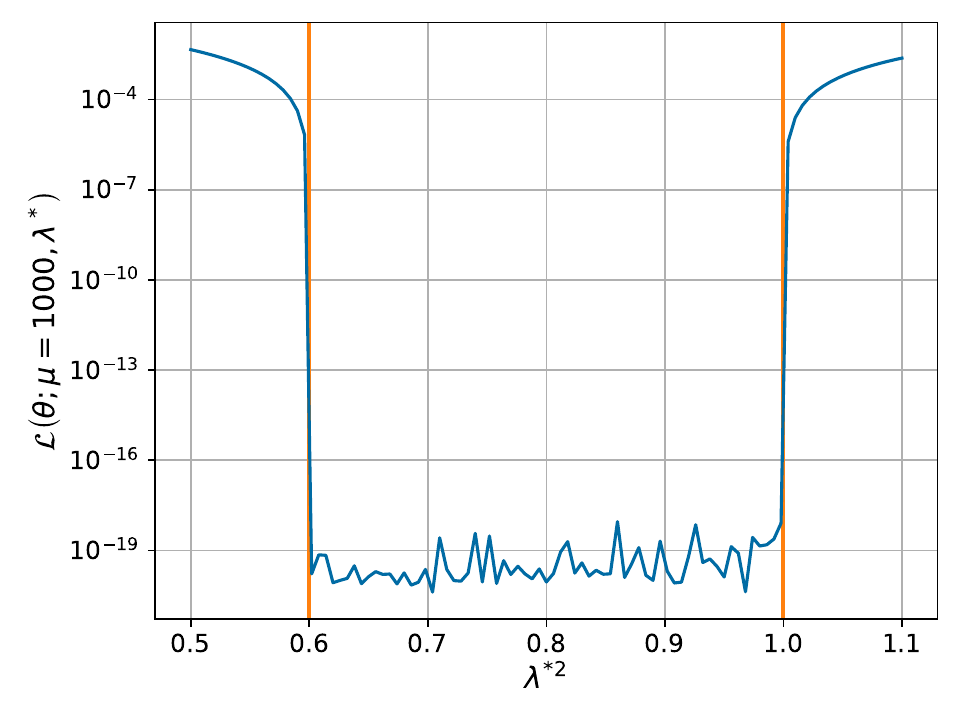}
    \caption{\label{fig:623-norm2}$\lambda^{*2}$ range for $((6,2,3))$ code. Penalty factor is chosen $\mu=1000$.}
\end{figure}

To find the range of $\lambda^*$, we sample $\lambda^*\in [0.5,1.1]$, then calculate the optimal value for $\mathcal{L}\left(\theta;\mu,\lambda^{*}\right)$ in eq(\ref{eq:lambda-length}). The results are shown in Fig \ref{fig:623-norm2}. For all optimizations, the violations of error-correcting conditions are less than $\mathcal{L}_{\mathrm{KL}}\leq 10^{-15}$. From the figure, a sharp transition from almost zero to nonzero can be observed, which indicates $\|\lambda\|_2^2\in [0.6,1.0]$. This two boundaries are also found via optimizing $\mathcal{L}_{\overline{\lambda}}$ and $\mathcal{L}_{\underline{\lambda}}$.

\subsection{Families of $((6,2,3))$ codes with $\sqrt{0.6}\leq \lambda^*\leq 1$}

To construct codes with $\sqrt{0.6}\leq \lambda^*\leq 1$, denote the six qubits by $q_1q_2q_3q_4q_5q_6$, and choose the following five bases for qubits
$q_2q_3q_4q_5q_6$
\begin{align*}
|S_{1}\rangle &= \frac{1}{\sqrt{2}} \left( \ket{00001} + \ket{11110} \right), \\
|S_{2}\rangle &= \frac{1}{\sqrt{2}} \left( \ket{00010} + \ket{11101} \right), \\
|S_{3}\rangle &= \frac{1}{\sqrt{2}} \left( \ket{00100} + \ket{11011} \right), \\
|S_{4}\rangle &= \frac{1}{\sqrt{2}} \left( \ket{01000} + \ket{10111} \right), \\
|S_{5}\rangle &= \frac{1}{\sqrt{2}} \left( \ket{10000} + \ket{01111} \right).
\end{align*}
Now choose logical states as:
\[ \left|0_{L}\right\rangle =\sum_{i=1}^{5}\left|x_{i}\right\rangle \left|S_{i}\right\rangle ,\;\left|1_{L}\right\rangle =\sum_{i=1}^{5}\left|y_{i}\right\rangle \left|S_{i}\right\rangle. \]
Here
\[ \left|x_{i}\right\rangle ={\gamma}_{i}\left|0\right\rangle +{\gamma}_{i+5}\left|1\right\rangle ,\;i=1,2,3,4,5 \]
\[ \left|y_{i}\right\rangle ={\gamma}_{i+5}^{*}\left|0\right\rangle -{\gamma}_{i}^{*}\left|1\right\rangle ,\;i=1,2,3,4,5 \]
The KL condition reduces to the following conditions on $\left|x_{i}\right\rangle$ and $\left|y_{i}\right\rangle$ (see Appendix~\ref{app:623} for details):
\[ \sum_{i}\left|x_{i}\right\rangle \left\langle x_{i}\right|=\sum_{i}\left|y_{i}\right\rangle \left\langle y_{i}\right|,\;\sum_{i}\left|y_{i}\right\rangle \left\langle x_{i}\right|=0. \]
Notice that this is equivalent to require that the RDM of $q_1$ is $\frac{I}{2}$.

Now let
\[
{\gamma}_j=a_j+ib_j,\ {\gamma}_{5+j}=c_j+id_j,
\]
for $j=1,2,3,4,5$. Where $a_j,b_j,c_j,d_j\in\mathbb{R}$.
Define the column vectors \( a \), \( b \), \( c \), and \( d \) to be:
\[
a = \begin{pmatrix}
a_1 \\
a_2 \\
a_3 \\
a_4 \\
a_5
\end{pmatrix}, \quad
b = \begin{pmatrix}
b_1 \\
b_2 \\
b_3 \\
b_4 \\
b_5
\end{pmatrix}, \quad
c = \begin{pmatrix}
c_1 \\
c_2 \\
c_3 \\
c_4 \\
c_5
\end{pmatrix}, \quad
d = \begin{pmatrix}
d_1 \\
d_2 \\
d_3 \\
d_4 \\
d_5
\end{pmatrix}.
\]
Then the KL condition will give the following conditions on $a,b,c,d$:
\begin{equation}
\begin{aligned}
a \cdot a &= b \cdot b = c \cdot c = d \cdot d = \frac{1}{4}, \\
a \cdot b &= a \cdot c = a \cdot d = b \cdot c = b \cdot d = c \cdot d = 0.
\end{aligned}
\end{equation}
This means that $a,b,c,d$ are orthogonal vectors in $\mathbb{R}^5$.

We then choose $e$ being the vector orthogonal to $a,b,c,d$, that is, the unnormalized orthogonal matrix composed from $(a,b,c,d,e)$ as $A$
\[ A=\left[\begin{array}{ccccc}
a & b & c & d & e\end{array}\right]=\left[\begin{array}{ccccc}
a_{1} & b_{1} & c_{1} & d_{1} & e_{1}\\
a_{2} & b_{2} & c_{2} & d_{2} & e_{2}\\
a_{3} & b_{3} & c_{3} & d_{3} & e_{3}\\
a_{4} & b_{4} & c_{4} & d_{4} & e_{4}\\
a_{5} & b_{5} & c_{5} & d_{5} & e_{5}
\end{array}\right]
\]
\[\;\rightarrow\;AA^{T}=A^{T}A=\frac{1}{4}I \]
which means each column (row) are orthogonal to each other.
In other words, $2A$ is a $5\times 5$ orthogonal matrix.

It turns out that the nonzero element of the signature vector, denoted as
$
PE^{\dag}_iE_jP=\lambda_{ij}P
$
of this code is given by the element of $e$ (see Appendix~\ref{app:623} for details):
\[
\begin{aligned}
\lambda_{X_{i}X_{j}} &= \lambda_{Y_{i}Y_{j}} = -2e_{7-i}e_{7-j}, \\
\lambda_{Z_{i}Z_{j}} &= 2e_{7-i}^{2} + 2e_{7-j}^{2}, \quad i,j \in \left\{ 2,3,4,5,6 \right\}.
\end{aligned}
\]
And
\[ \lambda^{*2}=\frac{1}{2}+8\sum_{i}e_{i}^{4}. \]

This means that $\lambda^{*}$ is invariant with the rotation within the subspace
spanned by $(a,b,c,d)$ . To further understand this invariance, we can view $\left|0_{L}\right\rangle$ and $\left|1_{L}\right\rangle$ as bipartite states
between $q_1$ (Party I) and $q_2q_3q_4q_5q_6$ (Party II). For Party I, an orthogonal transformation (i.e. $2A$, change of basis in the subspace spanned by $\left|S_{i}\right\rangle$) will correspond to a unitary transformation on Party II, hence will not change the RDM of Party I.  This unitary in general cannot be realized by LU transformations on $q_2q_3q_4q_5q_6$,  hence will lead to LU inequivalent codes.

It turns out (see Appendix~\ref{app:623} for details), however, when $e$ is chosen, the freedom in the choice of $(a,b,c,d)$ will lead to locally equivalent codes. This is due to the fact that, all such choices, given by
\[
\left[\begin{array}{cccc}
a & b & c & d \end{array}\right]O,
\]
where $O$ is any $4\times 4$ orthogonal matrix, can be generated by
\begin{itemize}
\item[1.] local unitary transformations on party I (i.e. the first qubit) (leading to LU equivalent code), and
\item[2.] unitary transformations in the logical space spanned by $\left|0_{L}\right\rangle$ and $\left|1_{L}\right\rangle$ (leading to the same code).
\end{itemize}
In other words, the choice of $e$ will in general lead to local inequivalent codes. 

Specifically, the vector $e$ for $\lambda^{*2}_{\text{min}}=0.6$ is
\[ e=\frac{1}{2\sqrt{5}}\left[\begin{array}{ccccc}
1 & 1 & 1 & 1 & 1\end{array}\right], \]
and for $\lambda^{*2}_{\text{max}}=1$
\[ e=\frac{1}{2}\left[\begin{array}{ccccc}
0 & 0 & 0 & 0 & 1\end{array}\right]. \]

\subsection{A single-parameter family}

To have a single-parameter family of codes that connect $\lambda^{*}_{\text{min}}$ to
$\lambda^{*2}_{\text{max}}=1$,
since $\lambda^*$ is only dependent on $e$, let us choose a single parameter family for $e$
\[
e=\frac{1}{2}[\frac{1}{2}\sin{\theta},\frac{1}{2}\sin{\theta},\frac{1}{2}\sin{\theta},\frac{1}{2}\sin{\theta},\cos{\theta}],
\]
for $\cos{\theta}\in[\frac{1}{\sqrt{5}},1]$.

Now we can choose the matrix $A$ as

\[
\mathbf{A} = \frac{1}{2}\begin{bmatrix}
\frac{1}{2} & \frac{1}{2} & \frac{1}{2} & \frac{1}{2}\cos(\theta) & \frac{1}{2}\sin(\theta) \\
\frac{1}{2} & -\frac{1}{2} & -\frac{1}{2} & \frac{1}{2}\cos(\theta) & \frac{1}{2}\sin(\theta) \\
-\frac{1}{2} & \frac{1}{2} & -\frac{1}{2} & \frac{1}{2}\cos(\theta) & \frac{1}{2}\sin(\theta) \\
-\frac{1}{2} & -\frac{1}{2} & \frac{1}{2} & \frac{1}{2}\cos(\theta) & \frac{1}{2}\sin(\theta) \\
0 & 0 & 0 & -\sin(\theta) & \cos(\theta) \\
\end{bmatrix}
\]

This gives a single parameter family of codes with the corresponding
\[
\lambda^{*2}=\frac{1}{2}+\frac{1}{2}\left(\frac{1}{4}\sin^4(\theta)+\cos^4(\theta)\right)
\]
runs continuously from $0.6$ to $1$.

For this family of codes, the matrix \( \lambda_{ij} \) will be block diagonal, and each block corresponding to
$X_iX_j,Y_iY_j,Z_iZ_j$ correlations, with the form (we only need to consider $i,j \in \left\{ 2,3,4,5,6 \right\}$:
\[
B=\begin{bmatrix}
1 & r & r & r & r \\
r & 1 & s & s & s \\
r & s & 1 & s & s \\
r & s & s & 1 & s \\
r & s & s & s & 1
\end{bmatrix}.
\]

The eigenvalues of the matrix are:
\[
\lambda_{B1} = \lambda_{B2} = \lambda_{B3} = 1 - s,
\]
\[
\lambda_{B4} = \frac{2 + 3s + \sqrt{9s^2 + 16r^2}}{2},
\]
\[
\lambda_{B5} = \frac{2 + 3s - \sqrt{9s^2 + 16r^2}}{2}.
\]

Notice that

\begin{align}
    \lambda_{X_{2}X_{j}} &= \lambda_{Y_{2}Y_{j}} = -2e_{5}e_{7-j} \nonumber \\
    &= -2\left(\frac{1}{2}\cos{\theta}\frac{1}{4}\sin{\theta}\right),  \quad j \in \left\{ 2,3,4,5,6 \right\} \nonumber \\
    \lambda_{X_{i}X_{j}} &= \lambda_{Y_{i}Y_{j}} = -2e_{7-i}e_{7-j} \nonumber \\
    &= -2\left(\frac{1}{4}\sin{\theta}\right)^2, \quad i,j \in \left\{ 3,4,5,6 \right\}.
\end{align}

So the $X_iX_j,Y_iY_j$ blocks are the same and correspond to
\[
r=-\frac{1}{4}\sin{\theta}\cos{\theta},\ s=-2\left(\frac{1}{4}\sin{\theta}\right)^2.
\]

And
\begin{align}
    \lambda_{Z_{2}Z_{j}} &= 2e_{5}^{2} + 2e_{7-j}^{2} \nonumber \\
    &= 2\left(\frac{1}{4}\sin{\theta}\right)^2 + 2\left(\frac{1}{2}\cos{\theta}\right)^2,  \quad j \in \left\{ 2,3,4,5,6 \right\} \nonumber \\
    \lambda_{Z_{i}Z_{j}} &= 2e_{7-i}^{2} + 2e_{7-j}^{2} \nonumber \\
    &= 4\left(\frac{1}{4}\sin{\theta}\right)^2, \quad i,j \in \left\{ 3,4,5,6 \right\}.
\end{align}

The $Z_iZ_j$ block corresponds to
\[
r=2\left(\frac{1}{4}\sin{\theta}\right)^2+2\left(\frac{1}{2}\cos{\theta}\right)^2,\ s=4\left(\frac{1}{4}\sin{\theta}\right)^2.
\]

So the matrix \( \lambda_{ij} \)  will be full rank for $\cos{\theta}\in[\frac{1}{\sqrt{5}},1)$, i.e.  \( \lambda^{*}\in [\sqrt{0.6},1) \). For
$\cos{\theta}=1$, i.e.  \( \lambda^{*}=1 \), we have $\lambda_{B5}=0$.

The enumerator is found to be:
\begin{equation}
\begin{aligned}
A^{((6,2,3))}&=1+(\frac{3}{16} \cos (2 \theta)+\frac{5}{64} \cos (4 \theta)+\frac{47}{64} )z^2\\
&+(-\frac{3}{16} \cos (2 \theta)-\frac{5}{64} \cos (4 \theta)+\frac{17}{64})z^3 \\
&+(-\frac{3}{16} \cos (2 \theta)-\frac{5}{64} \cos (4 \theta)+\frac{721}{64})z^4 \\
&+(\frac{3}{16} \cos (2 \theta)+\frac{5}{64} \cos (4 \theta)+\frac{1007}{64})z^5 +3z^6
\end{aligned},
\end{equation}
\begin{equation}
\begin{aligned}
B^{((6,2,3))}&=1+(\frac{3}{16} \cos(2 \theta)+\frac{5}{64} \cos(4 \theta)+\frac{47}{64})z^2\\
&+(\frac{3}{8} \cos(2 \theta)+\frac{5}{32} \cos(4 \theta)+\frac{751}{32})z^3 \\
&+(-\frac{3}{4} \cos (2 \theta)-\frac{5}{16}\cos (4 \theta)+\frac{577}{16})z^4 \\
&+(-\frac{3}{8} \cos (2 \theta)-\frac{5}{32} \cos (4 \theta)+\frac{1297}{32})z^5 \\
&+(\frac{9}{16} \cos (2 \theta)+\frac{15}{64} \cos (4 \theta)+\frac{1677}{64})z^6
\end{aligned}.
\end{equation}

When $\cos(\theta) = 1$, we have $\lambda^{*2}_{\text{max}} = 1$, and
\[
e = \frac{1}{2} \begin{bmatrix} 0 & 0 & 0 & 0 & 1 \end{bmatrix}.
\]
The code subspace, spanned by $(\ket{0_L}, \ket{1_L})$, resides within the ground state space of the Hamiltonian
\[
H = -2 Z_{2} \sum_{i \in \{3,4,5,6\}} Z_i + \frac{1}{2} \sum_{i \in \{3,4,5,6\}} \sum_{\substack{j \in \{3,4,5,6\} \\ j \neq i}} Z_{i} Z_{j},
\]
which is 16-dimensional degenerate, and is spanned by
\[ \left|000001\right\rangle ,\left|000010\right\rangle ,\left|000100\right\rangle ,\left|001000\right\rangle ,\]
\[\left|011110\right\rangle ,\left|011101\right\rangle ,\left|011011\right\rangle ,\left|010111\right\rangle ,  \]
\[ \left|100001\right\rangle ,\left|100010\right\rangle ,\left|100100\right\rangle ,\left|101000\right\rangle ,\]
\[\left|111110\right\rangle ,\left|111101\right\rangle ,\left|111011\right\rangle ,\left|110111\right\rangle.  \]
This implies that the signature vector $\vec{P}$ lies on the boundary of $W^{(1)}(\{O_{\alpha}\})$, where $\text{wt}(O_{\alpha}) = 1,2$.

\section{The $((7,2,3))$ codes}
\label{sec:7-2-3}

For the $((7,2,3))$ case, consider the Steane code with stabilizers
\[
\begin{array}{r@{\hspace{0.5em}}cccccccc}
g_1 & : &X & I & X & I & X & I & X \\
g_2 & : & I & X & X & I & I & X & X \\
g_3 & : & I & I & I & X & X & X & X\\
g_4 & : & Z & I & Z & I & Z & I & Z \\
g_5 & : & I & Z & Z & I & I & Z & Z \\
g_6 & : & I & I & I & Z & Z & Z & Z
\end{array}
\]

This code has a signature vector $\vec{\lambda}=0$, corresponding to $\lambda^*=0$. To find the maximum value of $\lambda^*$, we run our algorithm and observe a sharp transition at $\lambda^*=\sqrt{7}$, as shown in Fig. 2.

\begin{figure}[h]
    \centering
    \includegraphics[width=0.9\linewidth]{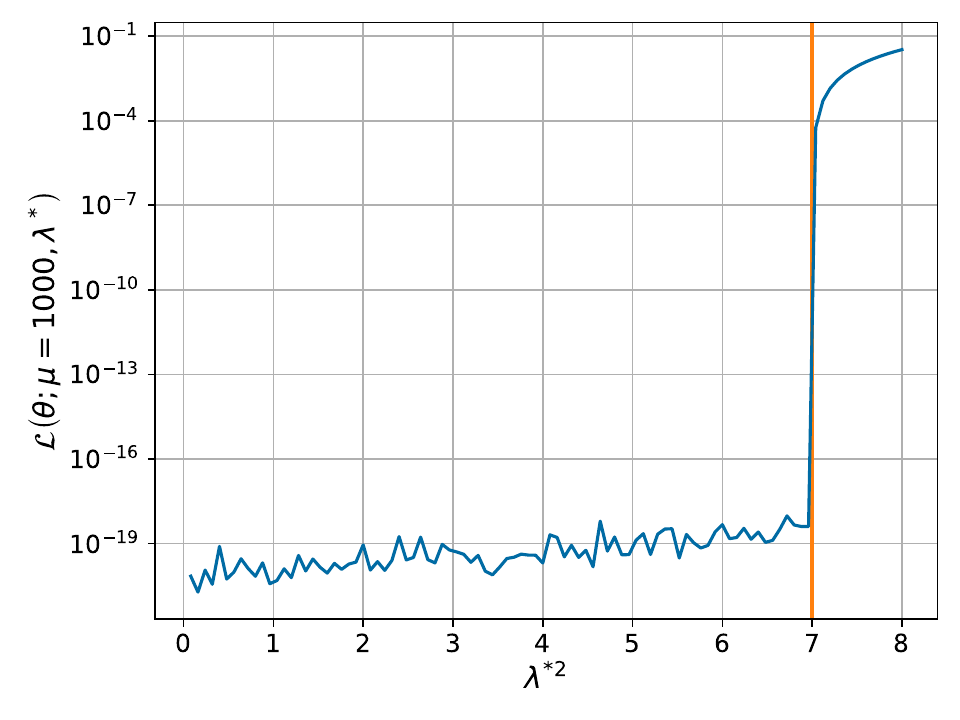}
    \caption{\label{fig:723-norm2} $\lambda^{*2}$ range for $((7,2,3))$ code.}
\end{figure}

It turns out that this maximum value $\lambda^*=\sqrt{7}$ corresponds to the permutation invariant code, which is constructed from the Dicke basis:
\[ D_{n,k}=\binom{n}{k}^{-1/2}\sum_{\sigma\in\mathrm{Sym}_{n}}\sigma\left|0\right\rangle ^{\otimes n-k}\otimes\left|1\right\rangle ^{\otimes k} \]
Two permutation invariant codes are given in~\cite{pollatsek2004permutationally} as:
\begin{equation}\label{eq:permutation-plus-qec}
\begin{aligned}
    8\left|0_{L}\right\rangle &= \sqrt{15}D_{7,0} - \sqrt{7}D_{7,2} + \sqrt{21}D_{7,4} + \sqrt{21}D_{7,6} \\
     |1_{L}\rangle &= X^{\otimes 7}|0_{L}\rangle
\end{aligned}
\end{equation}
and
\begin{equation}\label{eq:permutation-minus-qec}
\begin{aligned}
8\left|0_{L}\right\rangle &=\sqrt{15}D_{7,0}+\sqrt{7}D_{7,2}+\sqrt{21}D_{7,4}-\sqrt{21}D_{7,6}\\
    |1_{L}\rangle &= X^{\otimes 7}|0_{L}\rangle
\end{aligned}
\end{equation}

Notice that these two codes are local-unitary equivalent.

Now the key question is again, whether the set of all $\lambda^*$s is connected, i.e., whether the range of $\lambda^*$ is indeed $[0,\sqrt{7}]$. Notice that by permuting the qubits, Steane code can have cyclic symmetry, with logical $0$ and logical $1$ given by
\begin{equation}
\begin{aligned}
    \left|0_{L}\right\rangle &= \frac{1}{\sqrt{8}} \Big( \left|0000000\right\rangle + \left|1100101\right\rangle + \left|0101110\right\rangle + \left|0010111\right\rangle \\
    &\quad + \left|1001011\right\rangle + \left|1110010\right\rangle + \left|0111001\right\rangle + \left|1011100\right\rangle \Big) \\
    &= \frac{1}{\sqrt{8}}\left|0000000\right\rangle
    + \sqrt{\frac{7}{8}}\left(\left|0010111\right\rangle+\text{cyc.}\right)
\end{aligned}
\end{equation}
and $\left|1_{L}\right\rangle=X^{\otimes 7}\left|0_{L}\right\rangle.$
Here cyc. denotes all the other computational basis states with cyclic shift.
Now we will explicitly construct families of cyclic codes with $\lambda^*\in[0,\sqrt{7}]$.

Let us choose the cyclic basis with even weights:
\[ \left|\left\{ 0000000\right\} \right\rangle =\left|0000000\right\rangle\]
\[ \left|\left\{ 0000011\right\} \right\rangle =\frac{1}{\sqrt{7}}\left (|0000011\rangle + \text{cyc.} \right)\]
\[ \left|\left\{ 0000101\right\} \right\rangle =\frac{1}{\sqrt{7}}\left (|0000101\rangle + \text{cyc.} \right)\]
\[ \left|\left\{ 0001001\right\} \right\rangle =\frac{1}{\sqrt{7}}\left (|0001001\rangle + \text{cyc.} \right)\]
\[ \left|\left\{ 0001111\right\} \right\rangle =\frac{1}{\sqrt{7}}\left (|0001111\rangle + \text{cyc.} \right)\]
\[ \left|\left\{ 0011011\right\} \right\rangle =\frac{1}{\sqrt{7}}\left (|0011011\rangle + \text{cyc.} \right)\]
\[ \left|\left\{ 0011101\right\} \right\rangle =\frac{1}{\sqrt{7}}\left (|0011101\rangle + \text{cyc.} \right)\]
\[ \left|\left\{ 0101011\right\} \right\rangle =\frac{1}{\sqrt{7}}\left (|0101011\rangle + \text{cyc.} \right)\]
\[ \left|\left\{ 0010111\right\} \right\rangle =\frac{1}{\sqrt{7}}\left (|0010111\rangle + \text{cyc.} \right)\]
\[ \left|\left\{ 0111111\right\} \right\rangle =\frac{1}{\sqrt{7}}\left (|0111111\rangle + \text{cyc.} \right)\]

Using this basis, we parametrize $\left|0_{L}\right\rangle$ and $\left|1_L\right\rangle$ as follows:
\begin{equation}\label{eq:cyclic-code}
\begin{aligned}
    &\left|0_{L}\right\rangle = \ c_{0}\left|\{0000000\}\right\rangle\\
    &+ \frac{c_{1}}{\sqrt{3}} \Big( \left|\{0000011\}\right\rangle + \left|\{0000101\}\right\rangle + \left|\{0001001\}\right\rangle \Big)\\
    &+ c_{2}\left|\{0010111\}\right\rangle + \frac{c_{3}}{2} \Big( \left|\{0001111\}\right\rangle+ \left|\{0011011\}\right\rangle\\
    &+ \left|\{0011101\}\right\rangle + \left|\{0101011\}\right\rangle \Big)+ c_{4}\left|\{0111111\}\right\rangle, \\
    &\left|1_L\right\rangle = \ X^{\otimes 7}\left|0_L\right\rangle
\end{aligned}
\end{equation}

Within the five-dimensional subspace given in Eq(\ref{eq:cyclic-code}), KL conditions will lead to three independent equations. Combined with normalization condition, the coefficients $(c_0,c_1,c_2,c_3,c_4)$ should satisfy the following four equations:
\begin{gather}
    c_0^2 + c_1^2 + c_2^2 + c_3^2 + c_4^2 = 1 \label{eq:normalization} \\
    \begin{aligned}
    &\langle 0_{L}|Z_i|0_{L}\rangle = 0 \;\rightarrow\; \\
    &7c_{0}^{2} + 3c_{1}^{2} - c_{2}^{2} - c_{3}^{2} - 5c_{4}^{2} = 0
    \label{eq:0z0}
    \end{aligned} \\
    \begin{aligned}
    &\langle 0_{L}|X_i X^{\otimes7}|0_{L}\rangle  = 0 \;\rightarrow\; \\
    & 2\sqrt{7}c_{0}c_{4} + 2\sqrt{3}c_{1}c_{2} + 4\sqrt{3}c_{1}c_{3} \\
    & + 4\sqrt{3}c_{1}c_{4} + 4c_{2}c_{3} + 3c_{3}^{2} = 0
    \end{aligned} \label{eq:0x0} \\
    \begin{aligned}
    &\langle 0_{L}|Y_i X^{\otimes7}|0_{L}\rangle = 0 \;\rightarrow\; \\
    & 2\sqrt{7}c_{0}c_{4} + 2\sqrt{3}c_{1}c_{2} + 4\sqrt{3}c_{1}c_{3} \\
    & - 4\sqrt{3}c_{1}c_{4} - 4c_{2}c_{3} - 3c_{3}^{2} = 0
    \end{aligned} \label{eq:0y0}
\end{gather}

And for the signature vector, the following components are nonzero, satisfying (for $i\neq j$):

\begin{gather}
    21\langle0_{L}|X_{i}X_j|0_{L}\rangle = 2\sqrt{21}c_0c_1 + 10c_1^2 + 4\sqrt{3}c_1c_2 \notag \\
    + 8\sqrt{3}c_1c_3 + 12c_2c_3 + 6c_2c_4 + 9c_3^2 + 12c_3c_4 + 6c_4^2 \label{eq:0xx0} \\
    21\langle0_{L}|Y_{i}Y_j|0_{L}\rangle = -2\sqrt{21}c_0c_1 + 10c_1^2 - 4\sqrt{3}c_1c_2 \notag \\
    - 8\sqrt{3}c_1c_3 + 12c_2c_3 - 6c_2c_4 + 9c_3^2 - 12c_3c_4 + 6c_4^2 \label{eq:0yy0} \\
    21\langle0_{L}|Z_{i}Z_j|0_{L}\rangle = 21c_0^2 + c_1^2 - 3c_2^2 - 3c_3^2 + 9c_4^2 \label{eq:0zz0}
\end{gather}

From Eq.(\ref{eq:0x0}) and Eq.(\ref{eq:0y0}) we obtain:
\begin{gather}
    \sqrt{7}c_{0}c_{4}+2\sqrt{3}c_{1}c_{2}+2\sqrt{3}c_{1}c_{3}=0\label{eq:041213}\\
    4\sqrt{3}c_{1}c_{4}+4c_{2}c_{3}+3c_{3}^{2}=0\label{eq:142333}
\end{gather}

To solve these equations, we first find one solution for $c_4$ (see Appendix~\ref{app:723} for details):
\begin{equation}
    c_4=-\sqrt{3}c_1\label{eq:41linear}
\end{equation}

Then Eq.(\ref{eq:41linear}) and Eq.(\ref{eq:041213}) derive another linear relation:
\begin{equation}
    c_2 = -2c_3+\sqrt{7}c_0 \label{eq:230linear}
\end{equation}

Plug Eq.(\ref{eq:41linear}) and Eq.(\ref{eq:230linear}) into Eq.(\ref{eq:normalization}), (\ref{eq:0z0}), (\ref{eq:0xx0}), (\ref{eq:0yy0}), (\ref{eq:0zz0}), (\ref{eq:041213}) and (\ref{eq:142333}), one finds f (for $i\neq j$):\\
\begin{align}
&\text{Normalization: } \; \quad8 c_0^2-4 \sqrt{7} c_0 c_3+4 c_1^2+5 c_3^2=1\\
&\left\langle 0_L\left|Z_i\right| 0_L\right\rangle=0 \rightarrow \quad4 \sqrt{7} c_0 c_3-12 c_1^2-5 c_3^2=0 \label{eq:0z0mod}\\
&\left\langle 0_L\left|X_i X^{\otimes 7}\right| 0_L\right\rangle=0 \rightarrow \quad4 \sqrt{7} c_0 c_3-12 c_1^2-5 c_3^2=0 \\
&\left\langle 0_L\left|Y_i X^{\otimes 7}\right| 0_L\right\rangle=0 \rightarrow \quad-4 \sqrt{7} c_0 c_3+12 c_1^2+5 c_3^2=0 \\
&21\left\langle 0_L\left|X_i X_j\right| 0_L\right\rangle=\quad12 \sqrt{7} c_0 c_3+28 c_1^2-15 c_3^2\label{eq:0xx0mod}\\
&21\left\langle 0_L\left|Y_i Y_j\right| 0_L\right\rangle=\quad12 \sqrt{7} c_0 c_3+28 c_1^2-15 c_3^2\label{eq:0yy0mod}\\
&21\left\langle 0_L\left|Z_i Z_j\right| 0_L\right\rangle=\quad12 \sqrt{7} c_0 c_3+28 c_1^2-15 c_3^2\label{eq:0zz0mod}
\end{align}
Since the signature vector components ((\ref{eq:0xx0mod}), (\ref{eq:0yy0mod}) and (\ref{eq:0zz0mod})) are equal, it is convenient to introduce $\lambda^*$ as a parameter (for $i\ne j$):

\begin{align}
21\left\langle 0_L\left|X_i X_j\right| 0_L\right\rangle=\sqrt{7}\lambda^*\\
21\left\langle 0_L\left|Y_i Y_j\right| 0_L\right\rangle=\sqrt{7}\lambda^* \\
21\left\langle 0_L\left|Z_i Z_j\right| 0_L\right\rangle=\sqrt{7}\lambda^*
\label{eq:0zz0lambda}
\end{align}

By eliminating $c_0$ and $c_3$ through Eq.(\ref{eq:0z0mod}) and Eq.(\ref{eq:0xx0mod}), we find $c_{1}=\pm\frac{\sqrt{\sqrt{7}\lambda^{*}}}{8}$. With parameter $\lambda^*\in[0,\sqrt{7}]$, they become Steane code when $\lambda^*=0$, and parametric code at $\lambda^*=\sqrt{7}$. The following two QECCs are related to QECC in eq(\ref{eq:permutation-plus-qec}):
\begin{equation}
\begin{aligned}
    c_{0} &= \frac{\sqrt{\sqrt{7}\lambda^{*} + 8}}{8}, \\
    c_{1} &= -\frac{\sqrt{\sqrt{7}\lambda^{*}}}{8}, \\
    c_{4} &= -\sqrt{3}c_{1}, \\
    c_{3} &= \frac{2}{5} \left( \sqrt{7}c_{0} \pm \sqrt{7c_{0}^{2} - \frac{15\sqrt{7}\lambda^{*}}{64}} \right), \\
    c_{2} &= -2c_{3} + \sqrt{7}c_{0}
\end{aligned}\label{eq:723solution}
\end{equation}

The following two correspond to QECC in Eq.~(\ref{eq:permutation-minus-qec})
\begin{equation}\label{eq:coeff-cyclic-minus-code}
\begin{aligned}
    c_{0} &= \frac{\sqrt{\sqrt{7}\lambda^{*} + 8}}{8}, \\
    c_{1} &= \frac{\sqrt{\sqrt{7}\lambda^{*}}}{8}, \\
    c_{4} &= -\sqrt{3}c_{1}, \\
    c_{3} &= \frac{2}{5} \left( \sqrt{7}c_{0} \pm \sqrt{7c_{0}^{2} - \frac{15\sqrt{7}\lambda^{*}}{64}} \right), \\
    c_{2} &= -2c_{3} + \sqrt{7}c_{0}
\end{aligned}
\end{equation}
For the quantity inside the square root non-negative, it requires $\lambda^*\leq \sqrt{7}$. The signature vector for these four codes are the same with the following nonzero components (for $i\ne j$:
\[ \left\langle 0_{L}\right|X_{i}X_{j}\left|0_{L}\right\rangle =\left\langle 0_{L}\right|Y_{i}Y_{j}\left|0_{L}\right\rangle =\left\langle 0_{L}\right|Z_{i}Z_{j}\left|0_{L}\right\rangle =\frac{\lambda^{*}}{3\sqrt{7}}\]

This means that all the 2-particle reduced density matrix of the code have the form
\[
\rho^{(ij)}=\frac{1}{4}I+\frac{\lambda^{*}}{3\sqrt{7}}(X_iX_j+Y_iY_j+Z_iZ_j).
\]
Consequently, the matrix \( \lambda_{ij} \) will be block diagonal, and each block corresponding to
$X_iX_j,Y_iY_j,Z_iZ_j$ correlations, with the form
\[
(1 - s)I + sJ,\ s=\frac{\lambda^{*}}{3\sqrt{7}}\in[0,\frac{1}{3}].
\]
where:
\begin{itemize}
    \item \(I\) is the \(7 \times 7\) identity matrix.
    \item \(J\) is the \(7 \times 7\) matrix with all entries equal to \(1\).
\end{itemize}
This matrix $(1 - s)I + sJ$ has full rank and with one eigenvalues $6x + 1$ and six eigenvalues $1 - s$.

For the family given in Eq. (\ref{eq:723solution}), weight enumerators is given by
\begin{equation}
\begin{aligned}
A^{((7,2,3))}&= 1 + \lambda^{*2}z^2+ (21 - 2\lambda^{*2})z^4+ (42+\lambda^{*2})z^6\\
B^{((7,2,3))}&= 1 + \lambda^{*2}z^2+3(7 + \lambda^{*2})z^3+ (21-2\lambda^{*2})z^4\\
&+6(21-\lambda^{*2})z^5+(42+\lambda^{*2})z^6+3(15+\lambda^{*2})z^7.
\end{aligned}
\end{equation}

We have also explored all the local Clifford inequivalent \( ((7,2,3)) \) stabilizer codes, and found that the only possible values of \( \lambda^* \) are \( \{0, \sqrt{1}, \sqrt{2}, \sqrt{3}, \sqrt{5}\} \) (see Appendix~\ref{app:all723stab} for details). For instance, the Bare code~\cite{li2017fault} corresponds to \( \lambda^* = \sqrt{5} \).

When $\lambda^{*} = \lambda^{*}_{\text{max}} = \sqrt{7}$, the code subspace, spanned by $(\ket{0_L}, \ket{1_L})$, resides within the ground state space of the Hamiltonian
\[
H = -\sum_{i \neq j} \left( X_{i} X_{j} + Y_{i} Y_{j} + Z_{i} Z_{j} \right).
\]
This ground state space is 8-dimensional and corresponds to the symmetric subspace spanned by the Dicke basis. This implies that the signature vector $\vec{P}$ lies on the boundary of $W^{(1)}(\{O_{\alpha}\})$, where $\text{wt}(O_{\alpha}) = 1,2$.

\acknowledgments

We gratefully acknowledge Chenfeng Cao for valuable discussions.

\appendix

\section{Details of the ((6,2,3)) codes}
\label{app:623}

\subsection{The $|S_{i}\rangle$ basis}

In the six-qubits system $q_1q_2q_3q_4q_5q_6$, we choose a subspace for $q_2q_3q_4q_5q_6$ with basis $|S_i\rangle$:
\begin{align*}
|S_{1}\rangle &= \frac{1}{\sqrt{2}} \left( \ket{00001} + \ket{11110} \right), \\
|S_{2}\rangle &= \frac{1}{\sqrt{2}} \left( \ket{00010} + \ket{11101} \right), \\
|S_{3}\rangle &= \frac{1}{\sqrt{2}} \left( \ket{00100} + \ket{11011} \right), \\
|S_{4}\rangle &= \frac{1}{\sqrt{2}} \left( \ket{01000} + \ket{10111} \right), \\
|S_{5}\rangle &= \frac{1}{\sqrt{2}} \left( \ket{10000} + \ket{01111} \right).
\end{align*}

Their reduced density matrices (RDM) have clean forms. For example, 1-RDM
\[ 2\mathrm{Tr}_{q_{r}q_{s}q_{t}q_{\mu}}\left[\left|S_{i}\right\rangle \left\langle S_{j}\right|\right]=\delta_{ij}I_{2} \]
2-RDM are
\[ 4\mathrm{Tr}_{q_{r}q_{s}q_{t}}\left[\left|S_{i}\right\rangle \left\langle S_{i}\right|\right]=I_{4}+\left(-1\right)^{\delta}Z\otimes Z \]
where the sign $\delta\in\{0,1\}$ depends on how which qubits are chosen, for example:
\begin{align}
    4\mathrm{Tr}_{q_{2}q_{3}q_{4}} \left[\left|S_{1}\right\rangle \left\langle S_{1}\right|\right]
    &= I_{4} - Z \otimes Z, \nonumber \\
    4\mathrm{Tr}_{q_{2}q_{3}q_{6}} \left[\left|S_{1}\right\rangle \left\langle S_{1}\right|\right]
    &= I_{4} + Z \otimes Z.
\end{align}
Still, 2-RDM but with different basis
\[ 4\mathrm{Tr}_{q_{r}q_{s}q_{t}}\left[\left|S_{i}\right\rangle \left\langle S_{j}\right|\right]=\left(X\otimes X+Y\otimes Y\right)\delta^{(i,j)}, \]
\[ 4\mathrm{Tr}_{q_{2}q_{3}q_{4}}\left[\left|S_{1}\right\rangle \left\langle S_{2}\right|\right]=\left(X\otimes X+Y\otimes Y\right),\]
\[\mathrm{Tr}_{q_{2}q_{3}q_{5}}\left[\left|S_{1}\right\rangle \left\langle S_{2}\right|\right]=0. \]

Our code are designed in the following 10-dimensional subspace
\[ \left\{ \left|0\right\rangle ,\left|1\right\rangle \right\} \otimes\left\{ \left|S_{1}\right\rangle ,\left|S_{2}\right\rangle ,\left|S_{3}\right\rangle ,\left|S_{4}\right\rangle ,\left|S_{5}\right\rangle \right\}  \]
From above, all 2-RDM of pure states in this subspace will be in the form
\[ \mathrm{RDM}\left(q_{i},q_{j}\right)=\frac{1}{4}I_{4}+\alpha_{ij}\left(XX+YY\right)+\beta_{ij}ZZ. \]

The logical states are defined as
\[ \left|x_{i}\right\rangle =\gamma_{i}\left|0\right\rangle +\gamma_{i+5}\left|1\right\rangle ,\;i=1,2,3,4,5 \]
\[ \left|y_{i}\right\rangle =\gamma_{i+5}^{*}\left|0\right\rangle -\gamma_{i}^{*}\left|1\right\rangle ,\;i=1,2,3,4,5 \]
\[ \left|0_{L}\right\rangle =\sum_{i=1}^{5}\left|x_{i}\right\rangle \left|S_{i}\right\rangle ,\;\left|1_{L}\right\rangle =\sum_{i=1}^{5}\left|y_{i}\right\rangle \left|S_{i}\right\rangle. \]
It is convenient to introduce the following shorthand notation for the 2-RDMs:
\begin{align*}
    M_{ij}^{xx} &= \left\langle x_{i} \right| \left. x_{j} \right\rangle, \quad
    M_{ij}^{yx} = \left\langle y_{i} \right| \left. x_{j} \right\rangle, \nonumber \\
    M_{ij}^{xy} &= \left\langle x_{i} \right| \left. y_{j} \right\rangle, \quad
    M_{ij}^{yy} = \left\langle y_{i} \right| \left. y_{j} \right\rangle.
\end{align*}
\begin{align*}
    M\left(\square_{1}, \square_{2}, \square_{3}\right)
    &= \square_{1}\left(\left|00\right\rangle \left\langle 00\right|
    + \left|11\right\rangle \left\langle 11\right|\right) \nonumber \\
    &\quad + \square_{2}\left(\left|01\right\rangle \left\langle 01\right|
    + \left|10\right\rangle \left\langle 10\right|\right) \nonumber \\
    &\quad + \square_{3}\left(\left|01\right\rangle \left\langle 10\right|
    + \left|10\right\rangle \left\langle 01\right|\right).
\end{align*}
$M\left(\square_{1},\square_{2},\square_{3}\right)$ is a 4-by-4 matrix with diagonal elements $\square_1$ and $\square_2$, off-diagonal element $\square_3$. Four important properties about the tensor $M$ will be used later:
\begin{equation}\label{eq:M-property1}
    M_{ii}^{xy}=0
\end{equation}
\begin{equation} \label{eq:M-property2}
    M_{ii}^{xx}=M_{ii}^{yy}
\end{equation}
\begin{equation} \label{eq:M-property3}
    M_{ij}^{xx}=\gamma_{i}^{*}\gamma_{j}+\gamma_{i+5}^{*}\gamma_{j+5}=M_{ji}^{yy}
\end{equation}
\begin{align}
    M_{ij}^{xy} + M_{ji}^{xy} &= \left(\gamma_{i}^{*}\gamma_{j+5}^{*} - \gamma_{i+5}^{*}\gamma_{j}^{*}\right) \nonumber \\
    &\quad + \left(\gamma_{j}^{*}\gamma_{i+5}^{*} - \gamma_{j+5}^{*}\gamma_{i}^{*}\right) = 0.
    \label{eq:M-property4}
\end{align}

Since $\left\langle 0_{L}\right|\left.0_{L}\right\rangle=\sum_{i}\left\langle x_{i}\right|\left.x_{i}\right\rangle$, we have
\[ \left\langle 0_{L}\right|\left.0_{L}\right\rangle =\sum_{i}M_{ii}^{xx}=\sum_{i}\gamma_{i}\gamma_{i}^{*}=1. \]
With such a basis chosen, our code can already satisfy most of KL-conditions.

\subsection{The 2-RDMs and KL conditions}

\paragraph{RDM-$(q_2,q_3)$}
We first introduce RDM-$(q_2,q_3)$:
\begin{align}
    \mathrm{Tr}_{(23)^{c}}\left[\left|0_{L}\right\rangle \left\langle 0_{L}\right|\right]
    &= M\left(\square_{1}, \square_{2}, \square_{3}\right), \nonumber \\
    2\square_{1} &= M_{11}^{xx} + M_{22}^{xx} + M_{33}^{xx}, \nonumber \\
    2\square_{2} &= M_{44}^{xx} + M_{55}^{xx}, \nonumber \\
    2\square_{3} &= M_{45}^{xx} + M_{54}^{xx},
\end{align}

\begin{align}
    \mathrm{Tr}_{(23)^{c}}\left[\left|1_{L}\right\rangle \left\langle 1_{L}\right|\right]
    &= M\left(\square_{1}^{\prime}, \square_{2}^{\prime}, \square_{3}^{\prime}\right), \nonumber \\
    2\square_{1}^{\prime} &= M_{11}^{yy} + M_{22}^{yy} + M_{33}^{yy}, \nonumber \\
    2\square_{2}^{\prime} &= M_{44}^{yy} + M_{55}^{yy}, \nonumber \\
    2\square_{3}^{\prime} &= M_{45}^{yy} + M_{54}^{yy},
\end{align}

\begin{align}
    \mathrm{Tr}_{(23)^{c}}\left[\left|1_{L}\right\rangle \left\langle 0_{L}\right|\right]
    &= M\left(\square_{1}^{\prime\prime}, \square_{2}^{\prime\prime}, \square_{3}^{\prime\prime}\right), \nonumber \\
    2\square_{1}^{\prime\prime} &= M_{11}^{yx} + M_{22}^{yx} + M_{33}^{yx}, \nonumber \\
    2\square_{2}^{\prime\prime} &= M_{44}^{yx} + M_{55}^{yx}, \nonumber \\
    2\square_{3}^{\prime\prime} &= M_{45}^{yx} + M_{54}^{yx}.
\end{align}

\begin{itemize}
    \item From Eq.(\ref{eq:M-property2}), we have $\square_{1}=\square_{1}^{\prime}$ and $\square_{2}=\square_{2}^{\prime}$.
    \item From Eq.(\ref{eq:M-property3}), we have $\square_{3}=\square_{3}^{\prime}$.
    \item From Eq.(\ref{eq:M-property1}), we have $\square_{1}^{\prime\prime}=\square_{2}^{\prime\prime}=0$.
    \item From Eq.(\ref{eq:M-property4}), we have $\square_{3}^{\prime\prime}=0$
\end{itemize}
The $|0_L\rangle\langle 0_L|$ RDM is equal to $|1_L\rangle\langle 1_L|$ RDM and others are zero. It should be emphasized that this is true to all $\mathrm{RDM}(q_i,q_j),i\ne 1,j\ne 1$. \\

\paragraph{RDM-$(q_{i>1},q_{j>1})$} Without loss of generality, only $|0_L\rangle\langle 0_L|$ is listed below:
\begin{align}
    \mathrm{Tr}_{(24)^{c}}\left[\left|0_{L}\right\rangle \left\langle 0_{L}\right|\right]
    &= M\left(\square_{1}, \square_{2}, \square_{3}\right), \nonumber \\
    2\square_{1} &= M_{11}^{xx} + M_{22}^{xx} + M_{44}^{xx}, \nonumber \\
    2\square_{2} &= M_{33}^{xx} + M_{55}^{xx}, \nonumber \\
    2\square_{3} &= M_{35}^{xx} + M_{53}^{xx}, \nonumber
\end{align}

\begin{align}
    \mathrm{Tr}_{(25)^{c}}\left[\left|0_{L}\right\rangle \left\langle 0_{L}\right|\right]
    &= M\left(\square_{1}, \square_{2}, \square_{3}\right), \nonumber \\
    2\square_{1} &= M_{11}^{xx} + M_{33}^{xx} + M_{44}^{xx}, \nonumber \\
    2\square_{2} &= M_{22}^{xx} + M_{55}^{xx}, \nonumber \\
    2\square_{3} &= M_{25}^{xx} + M_{52}^{xx}, \nonumber
\end{align}

\begin{align}
    \mathrm{Tr}_{(26)^{c}}\left[\left|0_{L}\right\rangle \left\langle 0_{L}\right|\right]
    &= M\left(\square_{1}, \square_{2}, \square_{3}\right), \nonumber \\
    2\square_{1} &= M_{22}^{xx} + M_{33}^{xx} + M_{44}^{xx}, \nonumber \\
    2\square_{2} &= M_{11}^{xx} + M_{55}^{xx}, \nonumber \\
    2\square_{3} &= M_{15}^{xx} + M_{51}^{xx}, \nonumber
\end{align}

\begin{align}
    \mathrm{Tr}_{(34)^{c}}\left[\left|0_{L}\right\rangle \left\langle 0_{L}\right|\right]
    &= M\left(\square_{1}, \square_{2}, \square_{3}\right), \nonumber \\
    2\square_{1} &= M_{11}^{xx} + M_{22}^{xx} + M_{55}^{xx}, \nonumber \\
    2\square_{2} &= M_{33}^{xx} + M_{44}^{xx}, \nonumber \\
    2\square_{3} &= M_{34}^{xx} + M_{43}^{xx}, \nonumber
\end{align}

\begin{align}
    \mathrm{Tr}_{(35)^{c}}\left[\left|0_{L}\right\rangle \left\langle 0_{L}\right|\right]
    &= M\left(\square_{1}, \square_{2}, \square_{3}\right), \nonumber \\
    2\square_{1} &= M_{11}^{xx} + M_{33}^{xx} + M_{55}^{xx}, \nonumber \\
    2\square_{2} &= M_{22}^{xx} + M_{44}^{xx}, \nonumber \\
    2\square_{3} &= M_{24}^{xx} + M_{42}^{xx}, \nonumber
\end{align}

\begin{align}
    \mathrm{Tr}_{(36)^{c}}\left[\left|0_{L}\right\rangle \left\langle 0_{L}\right|\right]
    &= M\left(\square_{1}, \square_{2}, \square_{3}\right), \nonumber \\
    2\square_{1} &= M_{22}^{xx} + M_{33}^{xx} + M_{55}^{xx}, \nonumber \\
    2\square_{2} &= M_{11}^{xx} + M_{44}^{xx}, \nonumber \\
    2\square_{3} &= M_{14}^{xx} + M_{41}^{xx}, \nonumber
\end{align}

\begin{align}
    \mathrm{Tr}_{(45)^{c}}\left[\left|0_{L}\right\rangle \left\langle 0_{L}\right|\right]
    &= M\left(\square_{1}, \square_{2}, \square_{3}\right), \nonumber \\
    2\square_{1} &= M_{11}^{xx} + M_{44}^{xx} + M_{55}^{xx}, \nonumber \\
    2\square_{2} &= M_{22}^{xx} + M_{33}^{xx}, \nonumber \\
    2\square_{3} &= M_{23}^{xx} + M_{32}^{xx}, \nonumber
\end{align}

\begin{align}
    \mathrm{Tr}_{(46)^{c}}\left[\left|0_{L}\right\rangle \left\langle 0_{L}\right|\right]
    &= M\left(\square_{1}, \square_{2}, \square_{3}\right), \nonumber \\
    2\square_{1} &= M_{22}^{xx} + M_{44}^{xx} + M_{55}^{xx}, \nonumber \\
    2\square_{2} &= M_{11}^{xx} + M_{33}^{xx}, \nonumber \\
    2\square_{3} &= M_{13}^{xx} + M_{31}^{xx}, \nonumber
\end{align}

\begin{align}
    \mathrm{Tr}_{(56)^{c}}\left[\left|0_{L}\right\rangle \left\langle 0_{L}\right|\right]
    &= M\left(\square_{1}, \square_{2}, \square_{3}\right), \nonumber \\
    2\square_{1} &= M_{33}^{xx} + M_{44}^{xx} + M_{55}^{xx}, \nonumber \\
    2\square_{2} &= M_{11}^{xx} + M_{22}^{xx}, \nonumber \\
    2\square_{3} &= M_{12}^{xx} + M_{21}^{xx}.
\end{align}

\paragraph{RDM $(q_1,q_{k>1})$} To satisfy KL-conditions, the RDMs
\[ \mathrm{\mathrm{Tr}_{(1k)^{c}}}\left[\left|0_{L}\right\rangle \left\langle 0_{L}\right|\right]=\frac{1}{2}\sum_{i}\left|x_{i}\right\rangle \left\langle x_{i}\right|\otimes I_{2}, \]
\[ \mathrm{\mathrm{Tr}_{(1k)^{c}}}\left[\left|1_{L}\right\rangle \left\langle 1_{L}\right|\right]=\frac{1}{2}\sum_{i}\left|y_{i}\right\rangle \left\langle y_{i}\right|\otimes I_{2} \]
\[ \mathrm{\mathrm{Tr}_{(1k)^{c}}}\left[\left|1_{L}\right\rangle \left\langle 0_{L}\right|\right]=\frac{1}{2}\sum_{i}\left|y_{i}\right\rangle \left\langle x_{i}\right|\otimes I_{2} \]
must obey
\[ \sum_{i}\left|x_{i}\right\rangle \left\langle x_{i}\right|=\sum_{i}\left|y_{i}\right\rangle \left\langle y_{i}\right|,\;\sum_{i}\left|y_{i}\right\rangle \left\langle x_{i}\right|=0. \]

From above, we can derive
\[ \gamma=\left[\begin{array}{cc}
a+ib & c+id\end{array}\right]\in\mathbb{C}^{10},\;a,b,c,d\in\mathbb{R}^{5}, \]
\begin{align*}
    a \cdot a &= b \cdot b = c \cdot c = d \cdot d = \frac{1}{4}, \nonumber \\
    a \cdot b &= a \cdot c = a \cdot d = b \cdot c = b \cdot d = c \cdot d = 0.
\end{align*}
\[ \sum_{i}\left|x_{i}\right\rangle \left\langle x_{i}\right|=\frac{1}{2}I_2\]
Thus, vector $(a,b,c,d)$ make four columns of orthogonal group $O(5)$.

\subsection{Expression of $\lambda^*$}

Let $e$ being the vector orthogonal to $a,b,c,d$
\[ e\cdot e=\frac{1}{4},e\cdot a=e\cdot b=e\cdot c=e\cdot d=0,e\in\mathbb{R}^{5} \]
Then we can prove
\[ M_{ii}^{xx}=\frac{1}{4}-e_{i}^{2},M_{ij}^{xx}+M_{ji}^{xx}=-2e_{i}e_{j}. \]
\begin{proof} Let's take $i=1$ for example. Denote the unnormalized orthogonal matrix composed from $(a,b,c,d,e)$ as:
\begin{align}
    A &= \left[\begin{array}{ccccc}
    a & b & c & d & e
    \end{array}\right] \nonumber \\
    &= \left[\begin{array}{ccccc}
    a_{1} & b_{1} & c_{1} & d_{1} & e_{1}\\
    a_{2} & b_{2} & c_{2} & d_{2} & e_{2}\\
    a_{3} & b_{3} & c_{3} & d_{3} & e_{3}\\
    a_{4} & b_{4} & c_{4} & d_{4} & e_{4}\\
    a_{5} & b_{5} & c_{5} & d_{5} & e_{5}
    \end{array}\right]
\end{align}
\[    \rightarrow  AA^{T} = A^{T}A = \frac{1}{4}I.\]
which means each column (row) are orthogonal to each other. Then
\[ M_{11}^{xx}=a_{1}^{2}+b_{1}^{2}+c_{1}^{2}+d_{1}^{2}=\frac{1}{4}-e_{1}^{2}, \]
and similarly
\begin{align}
    M_{12}^{xx} + M_{21}^{xx} &= a_{1}a_{2} + b_{1}b_{2} + c_{1}c_{2} + d_{1}d_{2} \nonumber \\
    &= \left(a_{1}a_{2} + b_{1}b_{2} + c_{1}c_{2} + d_{1}d_{2} + e_{1}e_{2}\right) \nonumber \\
    &\quad - e_{1}e_{2} \nonumber \\
    &= - e_{1}e_{2}.
\end{align}
\end{proof}
All nonzero components of signature vector can be written as
\begin{align}
    \lambda_{X_{i}X_{j}} &= \lambda_{Y_{i}Y_{j}} = -2e_{7-i}e_{7-j}, \nonumber \\
    \lambda_{Z_{i}Z_{j}} &= 2e_{7-i}^{2} + 2e_{7-j}^{2}, \quad i,j \in \left\{ 2,3,4,5,6 \right\}.
\end{align}
For example
\[ \lambda_{X_{2}X_{3}}=\lambda_{Y_{2}Y_{3}}=-2e_{4}e_{5},\lambda_{Z_{2}Z_{3}}=2e_{4}^{2}+2e_{5}^{2}.\]
Then its square of length is
\[ \lambda^{*2}=\frac{1}{2}+8\sum_{i}e_{i}^{4} \]
\begin{proof} Two lemmas:
    \[ \sum_{i}\sum_{j\ne i}e_{i}^{2}e_{j}^{2}=\left(\sum_{i}e_{i}^{2}\right)^{2}-\sum_{i}e_{i}^{4}=\frac{1}{16}-\sum_{i}e_{i}^{4} \]
    \[ \sum_{i}\sum_{j\ne i}e_{i}^{4}=\sum_{i}\sum_{j}e_{i}^{4}-\sum_{i}e_{i}^{4}=4\sum_{i}e_{i}^{4} \]
    Using lemmas above,
    \begin{align*}
        \lambda^{*2}&=\sum_{i}\sum_{j>i}4e_{i}^{2}e_{j}^{2}+4e_{i}^{2}e_{j}^{2}+4\left(e_{i}^{2}+e_{j}^{2}\right)^{2}\\
        &=\sum_{i}\sum_{j\ne i}4e_{i}^{2}e_{j}^{2}+2\left(e_{i}^{2}+e_{j}^{2}\right)^{2}\\
        &=\sum_{i}\sum_{j\ne i}8e_{i}^{2}e_{j}^{2}+2e_{i}^{4}+2e_{j}^{4}\\
        &=\frac{1}{2}+8\sum_{i}e_{i}^{4}
    \end{align*}
\end{proof}

\subsection{Structure of SO(4) symmetry}

The Lie algebra \(\mathfrak{so}(4)\) comprises all \(4 \times 4\) skew-symmetric matrices. A general skew-symmetric matrix \(X \in \mathfrak{so}(4)\) satisfies $X^T = -X$.
Such matrices have zeros on the diagonal and contain six independent off-diagonal elements. These non-zero elements correspond to infinitesimal rotations in the six independent planes of four-dimensional space ($x_i-x_j$ plane for $i\neq j$). Each generator can be represented by a matrix \(E_{ij}\), where the \((i,j)\)-th entry is \(+1\), the \((j,i)\)-th entry is \(-1\), and all other entries are zero.

To obtain the Lie group \(\mathrm{SO}(4)\), we exponentiate the Lie algebra elements. Specifically, for any skew-symmetric matrix \(X \in \mathfrak{so}(4)\), the corresponding element in \(\mathrm{SO}(4)\) can be obtained as:

\[
\mathrm{SO}(4) = \left\{ e^{X} \mid X \in \mathfrak{so}(4) \right\}
\]

Now choose the set
\[
K_1=E_{12}+E_{34}, K_2=E_{12}-E_{34},
\]
\[
K_3=E_{23}+E_{14}, K_4=E_{23}-E_{14},
\]
\[
K_5=E_{13}+E_{24}, K_6=E_{13}-E_{24}.
\]

Notice that

\[
\exp(\theta K_1) = \begin{pmatrix}
\cos\theta & \sin\theta & 0 & 0 \\
-\sin\theta & \cos\theta & 0 & 0 \\
0 & 0 & \cos\theta & \sin\theta \\
0 & 0 & -\sin\theta & \cos\theta
\end{pmatrix}
\]

\[
\exp(\theta K_2) = \begin{pmatrix}
\cos\theta & \sin\theta & 0 & 0 \\
-\sin\theta & \cos\theta & 0 & 0 \\
0 & 0 & \cos\theta & -\sin\theta \\
0 & 0 & \sin\theta & \cos\theta
\end{pmatrix}
\]

\[
\exp(\theta K_3) = \begin{pmatrix}
\cos\theta & 0 & 0 & \sin\theta \\
0 & \cos\theta & \sin\theta & 0 \\
0 & -\sin\theta & \cos\theta & 0 \\
-\sin\theta & 0 & 0 & \cos\theta
\end{pmatrix}
\]

\[
\exp(\theta K_4) = \begin{pmatrix}
\cos\theta & 0 & 0 & -\sin\theta \\
0 & \cos\theta & \sin\theta & 0 \\
0 & -\sin\theta & \cos\theta & 0 \\
\sin\theta & 0 & 0 & \cos\theta
\end{pmatrix}
\]

\[
\exp(\theta K_5) = \begin{pmatrix}
\cos\theta & 0 & \sin\theta & 0 \\
0 & \cos\theta & 0 & \sin\theta \\
-\sin\theta & 0 & \cos\theta & 0 \\
0 & -\sin\theta & 0 & \cos\theta
\end{pmatrix}
\]

\[
\exp(\theta K_6) = \begin{pmatrix}
\cos\theta & 0 & \sin\theta & 0 \\
0 & \cos\theta & 0 & -\sin\theta \\
-\sin\theta & 0 & \cos\theta & 0 \\
0 & \sin\theta & 0 & \cos\theta
\end{pmatrix}
\]

Notice that

\[
\exp(-i{\theta X})=\cos{\theta}I-i\sin{\theta}X=
\begin{bmatrix} \cos{\theta} & -i\sin{\theta} \\
-i\sin{\theta} & \cos{\theta} \end{bmatrix}
\]
\[
\exp(-i{\theta Y})=\cos{\theta}I-i\sin{\theta}Y=
\begin{bmatrix} \cos{\theta} & -\sin{\theta} \\
\sin{\theta}& \cos{\theta} \end{bmatrix}
\]
\[
\exp(-i{\theta Z})=\cos{\theta}I-i\sin{\theta}Z=
\begin{bmatrix} e^{-i\theta} & 0 \\ 0 & e^{i\theta} \end{bmatrix}.
\]

Write
\[
\ket{x_i}=(a_i+ib_i)\ket{0}+(c_i+id_i)\ket{1}=\begin{bmatrix} a_i+ib_i \\ c_i+id_i \end{bmatrix}.
\]
\[
\ket{y_i}=(c_i-id_i)\ket{0}-(a_i-ib_i)\ket{1}=\begin{bmatrix} c_i-id_i \\ -a_i+ib_i\end{bmatrix}.
\]

We now compare
\[
\exp{\theta  K_i }
\]\\

with the representation of
\begin{itemize}
    \item $\exp{i\theta X_1}, \exp{i\theta Y_1}, \exp{i\theta Z_1}$
    \item $\exp{i\theta X_L}, \exp{i\theta Y_L}, \exp{i\theta Z_L}$
\end{itemize}

as \(\mathrm{SO}(4)\) rotations\\

\paragraph{$\exp{(-i\theta X_1)}$}
\[
\exp{(-i\theta X_1)}\ket{x_i}=
\begin{bmatrix} \cos{\theta} & -i\sin{\theta} \\
-i\sin{\theta} & \cos{\theta} \end{bmatrix}
\begin{bmatrix} a_i+ib_i \\ c_i+id_i \end{bmatrix}
\]
\[
=\begin{bmatrix} (\cos{\theta}a_i+\sin{\theta}d_i)+ i(\cos{\theta}b_i-\sin{\theta}c_i)\\
(\sin{\theta}b_i+\cos{\theta}c_i)+ i(-\sin{\theta}a_i+\cos{\theta}d_i)
\end{bmatrix}
\]

As a \(\mathrm{SO}(4)\) rotation, this is
\[\left[\begin{array}{cccc}
a & b & c & d\end{array}\right]
\left[\begin{array}{cccc}
\cos{\theta} & 0 & 0 & -\sin{\theta}\\
 0 & \cos{\theta} & \sin{\theta}  & 0\\
0 &-\sin{\theta}&\cos{\theta} & 0\\
\sin{\theta} & 0 & 0 & \cos{\theta}
\end{array}\right],\]\\

which corresponds to \(\exp(\theta K_4)\).\\

\paragraph{$\exp{(-i\theta Y_1)}$}
\[
\exp{(-i\theta Y_1)}\ket{x_i}=
\begin{bmatrix} \cos{\theta} & -\sin{\theta} \\
\sin{\theta} & \cos{\theta} \end{bmatrix}
\begin{bmatrix} a_i+ib_i \\ c_i+id_i \end{bmatrix}
\]
\[
=\begin{bmatrix} (\cos{\theta}a_i-\sin{\theta}c_i)+ i(\cos{\theta}b_i-\sin{\theta}d_i)\\
(\sin{\theta}a_i+\cos{\theta}c_i)+ i(\sin{\theta}b_i+\cos{\theta}d_i)
\end{bmatrix}
\]\\

As a \(\mathrm{SO}(4)\) rotation, this is
\[\left[\begin{array}{cccc}
a & b & c & d\end{array}\right]
\left[\begin{array}{cccc}
\cos{\theta} & 0 & \sin{\theta} & 0\\
 0 & \cos{\theta} & 0 & \sin{\theta} \\
-\sin{\theta}&0 & \cos{\theta} & 0\\
0& -\sin{\theta} & 0 & \cos{\theta}
\end{array}\right],
\]\\

which corresponds to \(\exp(\theta K_5)\).\\

\paragraph{$\exp{(-i\theta Z_1)}$}
\[
\exp{(-i\theta Z_1)}\ket{x_i}=\]
\[
\begin{bmatrix} \cos{\theta}-i \sin{\theta} & 0\\
0 & \cos{\theta}+i\sin{\theta} \end{bmatrix}
\begin{bmatrix} a_i+ib_i \\ c_i+id_i \end{bmatrix}
\]
\[
=\begin{bmatrix} (\cos{\theta}a_i+\sin{\theta}b_i)+ i(-\sin{\theta}a_i+\cos{\theta}b_i)\\
(\cos{\theta}c_i-\sin{\theta}d_i+)+ i(\sin{\theta}c_i+\cos{\theta}d_i)
\end{bmatrix}
\]\\

As a \(\mathrm{SO}(4)\) rotation, this is
\[\left[\begin{array}{cccc}
a & b & c & d\end{array}\right]
\left[\begin{array}{cccc}
\cos{\theta} & -\sin{\theta} & 0 & 0\\
 \sin{\theta} & \cos{\theta} & 0 & 0  \\
0 &0 & \cos{\theta} & \sin{\theta}\\
0 & 0& -\sin{\theta}  & \cos{\theta}
\end{array}\right],
\]\\

which corresponds to \(\exp(-\theta K_2)\).\\

\paragraph{$\exp{(-i\theta X_L)}$}
\[
\exp{(-i\theta X_L)}\ket{0_L}=\cos{\theta}\ket{0_L}-i\sin{\theta}\ket{1_L}
\]
\[
\rightarrow\cos{\theta}\begin{bmatrix} a_i+ib_i \\ c_i+id_i \end{bmatrix}
-i\sin{\theta}\begin{bmatrix} c_i-id_i \\ -a_i+ib_i\end{bmatrix}
\]
\[
=\begin{bmatrix} (\cos{\theta}a_i-\sin{\theta}d_i)+ i(\cos{\theta}b_i-\sin{\theta}c_i)\\
(\sin{\theta}b_i+\cos{\theta}c_i)+ i(\sin{\theta}a_i+\cos{\theta}d_i)
\end{bmatrix}
\]\\

As a \(\mathrm{SO}(4)\) rotation, this is
\[\left[\begin{array}{cccc}
a & b & c & d\end{array}\right]
\left[\begin{array}{cccc}
\cos{\theta} & 0 & 0 & \sin{\theta}\\
 0 & \cos{\theta} & \sin{\theta}  & 0\\
0 &-\sin{\theta}&\cos{\theta} & 0\\
-\sin{\theta} & 0 & 0 & \cos{\theta}
\end{array}\right],\]\\

which corresponds to \(\exp(\theta K_3)\).\\

\paragraph{$\exp{(-i\theta Y_L)}$}
\[
\exp{(-i\theta Y_L)}\ket{0_L}=\cos{\theta}\ket{0_L}+\sin{\theta}\ket{1_L}
\]
\[
\rightarrow\cos{\theta}\begin{bmatrix} a_i+ib_i \\ c_i+id_i \end{bmatrix}
+\sin{\theta}\begin{bmatrix} c_i-id_i \\ -a_i+ib_i\end{bmatrix}
\]
\[
=\begin{bmatrix} (\cos{\theta}a_i+\sin{\theta}c_i)+ i(\cos{\theta}b_i-\sin{\theta}d_i)\\
(-\sin{\theta}a_i+\cos{\theta}c_i)+ i(\sin{\theta}b_i+\cos{\theta}d_i)
\end{bmatrix}
\]\\

As a \(\mathrm{SO}(4)\) rotation, this is
\[\left[\begin{array}{cccc}
a & b & c & d\end{array}\right]
\left[\begin{array}{cccc}
\cos{\theta} & 0 & -\sin{\theta} & 0\\
 0 & \cos{\theta} & 0 & \sin{\theta} \\
\sin{\theta}&0 & \cos{\theta} & 0\\
0& -\sin{\theta} & 0 & \cos{\theta}
\end{array}\right],
\]\\

which corresponds to \(\exp(-\theta K_6)\).\\

\paragraph{$\exp{(-i\theta Z_L)}$}
\[
\exp{(-i\theta Z_L)}\ket{0_L}=(\cos{\theta}-i\sin{\theta})\ket{0_L}\]
\[
\rightarrow(\cos{\theta}-i\sin{\theta})\begin{bmatrix} a_i+ib_i \\ c_i+id_i \end{bmatrix}
\]
\[
=\begin{bmatrix} (\cos{\theta}a_i+\sin{\theta}b_i)+ i(-\sin{\theta}a_i+\cos{\theta}b_i)\\
(\cos{\theta}c_i+\sin{\theta}d_i+)+ i(-\sin{\theta}c_i+\cos{\theta}d_i)
\end{bmatrix}
\]\\

As a \(\mathrm{SO}(4)\) rotation, this is
\[\left[\begin{array}{cccc}
a & b & c & d\end{array}\right]
\left[\begin{array}{cccc}
\cos{\theta} & -\sin{\theta} & 0 & 0\\
 \sin{\theta} & \cos{\theta} & 0 & 0  \\
0 &0 & \cos{\theta} & -\sin{\theta}\\
0 & 0& \sin{\theta}  & \cos{\theta}
\end{array}\right],
\]\\

which corresponds to \(\exp(-\theta K_1)\).

\section{Details for the ((7,2,3)) codes}
\label{app:723}

We provide the details for solving the following
system of equations:

\begin{align}
    c_0^2 + c_1^2 + c_2^2 + c_3^2 + c_4^2 &= 1 \tag{E1} \label{E1} \\
    7c_0^2 + 3c_1^2 - c_2^2 - c_3^2 - 5c_4^2 &= 0 \tag{E2} \label{E2} \\
    2\sqrt{7}\, c_0 c_4 + 2\sqrt{3}\, c_1 c_2 + 4\sqrt{3}\, c_1 c_3 & \notag \\
    + 4\sqrt{3}\, c_1 c_4+ 4 c_2 c_3 + 3 c_3^2 &= 0 \tag{E3} \label{E3} \\
    2\sqrt{7}\, c_0 c_4 + 2\sqrt{3}\, c_1 c_2 + 4\sqrt{3}\, c_1 c_3 &\notag \\
    - 4\sqrt{3}\, c_1 c_4 - 4 c_2 c_3 - 3 c_3^2 &= 0 \tag{E4} \label{E4}
\end{align}

Our goal is to eliminate \( c_0 \), \( c_2 \), and \( c_3 \) to obtain an equation in terms of \( c_1 \) and \( c_4 \).

\subsection*{Step 1: Subtract Eq.(E4) from Eq.(E3)}

Subtracting Eq.\eqref{E4} from Eq.\eqref{E3}:

\begin{align*}
    & \left[ 2\sqrt{7}\, c_0c_4
          + 2\sqrt{3}\, c_1 c_2
          + 4\sqrt{3}\, c_1 c_3 \right. \nonumber \\
    & \quad \left. + 4\sqrt{3}\, c_1 c_4
          + 4 c_2 c_3
          + 3 c_3^2 \right] \nonumber \\
    & - \left[ 2\sqrt{7}\, c_0 c_4
          + 2\sqrt{3}\, c_1 c_2
          + 4\sqrt{3}\, c_1 c_3 \right. \nonumber \\
    & \quad \left. - 4\sqrt{3}\, c_1 c_4
          - 4 c_2 c_3
          - 3 c_3^2 \right] = 0
\end{align*}

Simplify:

\begin{align*}
    8\sqrt{3}\, c_1 c_4 + 8 c_2 c_3 + 6 c_3^2 &= 0 \\
    \Rightarrow \quad 4\sqrt{3}\, c_1 c_4 + 4 c_2 c_3 + 3 c_3^2 &= 0 \tag{E5} \label{E5}
\end{align*}

\subsection*{Step 2: Add Equations (E3) and (E4)}

Adding Equations \eqref{E3} and \eqref{E4}:

\begin{align*}
    & \Big[ 2\sqrt{7}\, c_0 c_4 + 2\sqrt{3}\, c_1 c_2 + 4\sqrt{3}\, c_1 c_3 \nonumber \\
    & + 4\sqrt{3}\, c_1 c_4 + 4 c_2 c_3 + 3 c_3^2 \Big] \nonumber \\
    & + \Big[ 2\sqrt{7}\, c_0 c_4 + 2\sqrt{3}\, c_1 c_2 + 4\sqrt{3}\, c_1 c_3\nonumber \\
    &  - 4\sqrt{3}\, c_1 c_4 - 4 c_2 c_3 - 3 c_3^2 \Big] = 0
\end{align*}

Simplify:

\begin{align*}
    4\sqrt{7}\, c_0 c_4 + 4\sqrt{3}\, c_1 c_2 + 8\sqrt{3}\, c_1 c_3 &= 0 \\
    \Rightarrow \quad \sqrt{7}\, c_0 c_4 + \sqrt{3}\, c_1 c_2 + 2\sqrt{3}\, c_1 c_3 &= 0 \tag{E6} \label{E6}
\end{align*}

\subsection*{Step 3: Eliminate \( c_0 \) Using Equations (E1) and (E2)}

From Eq.\eqref{E1}:

\begin{equation}
    c_0^2 = 1 - c_1^2 - c_2^2 - c_3^2 - c_4^2 \tag{E1a} \label{E1a}
\end{equation}

Substitute \( c_0^2 \) into Eq.\eqref{E2}:

\begin{align*}
    7(1 - c_1^2 - c_2^2 - c_3^2 - c_4^2) + 3c_1^2 - c_2^2 - c_3^2 - 5c_4^2 &= 0 \\
    7 - 7c_1^2 - 7c_2^2 - 7c_3^2 - 7c_4^2 + 3c_1^2 - c_2^2 - c_3^2 - 5c_4^2 &= 0
\end{align*}

Simplify:

\begin{equation}
    -4 c_1^2 - 8 c_2^2 - 8 c_3^2 - 12 c_4^2 + 7 = 0 \tag{E7} \label{E7}
\end{equation}

Divide both sides by $-1$:

\begin{equation}
    4 c_1^2 + 8 c_2^2 + 8 c_3^2 + 12 c_4^2 = 7 \tag{E7a} \label{E7a}
\end{equation}

Divide both sides by $4$:

\begin{equation}
    c_1^2 + 2 c_2^2 + 2 c_3^2 + 3 c_4^2 = \frac{7}{4} \tag{E8} \label{E8}
\end{equation}

\subsection*{Step 4: Use Eq.(E6) to Express \( c_0 c_4 \)}

From Eq.\eqref{E6}:

\begin{equation}
    \sqrt{7}\, c_0 c_4 = -\sqrt{3}\, c_1 c_2 - 2\sqrt{3}\, c_1 c_3 \tag{E9} \label{E9}
\end{equation}

\subsection*{Step 5: Compute \( (c_0 c_4)^2 \) from Eq.(E9)}

Square both sides of Eq.\eqref{E9}:

\begin{align*}
    (\sqrt{7}\, c_0 c_4)^2 &= \left( -\sqrt{3}\, c_1 c_2 - 2\sqrt{3}\, c_1 c_3 \right )^2 \\
    7 c_0^2 c_4^2 &= 3 c_1^2 \left( c_2 + 2 c_3 \right )^2 \\
    \Rightarrow \quad c_0^2 c_4^2 &= \frac{3}{7} c_1^2 \left( c_2 + 2 c_3 \right )^2 \tag{E10} \label{E10}
\end{align*}

\subsection*{Step 6: Express \( c_0^2 c_4^2 \) in Terms of \( c_1 \) and \( c_4 \)}

From Eq.\eqref{E1a}, we have:

\begin{equation}
    c_0^2 = 1 - c_1^2 - c_2^2 - c_3^2 - c_4^2 \tag{E1a}
\end{equation}

Therefore:

\begin{equation}
    c_0^2 c_4^2 = (1 - c_1^2 - c_2^2 - c_3^2 - c_4^2) c_4^2 \tag{E11} \label{E11}
\end{equation}

\subsection*{Step 7: Equate the Two Expressions for \( c_0^2 c_4^2 \)}

Set Eq.\eqref{E10} equal to Eq.\eqref{E11}:

\begin{equation}
    (1 - c_1^2 - c_2^2 - c_3^2 - c_4^2) c_4^2 = \frac{3}{7} c_1^2 \left( c_2 + 2 c_3 \right )^2 \tag{E12} \label{E12}
\end{equation}

\subsection*{Step 8: Use Eq.(E8) to Express \( c_2^2 + c_3^2 \)}

From Eq.\eqref{E8}:

\begin{equation}
    c_2^2 + c_3^2 = \frac{1}{2} \left( \frac{7}{4} - c_1^2 - 3 c_4^2 \right ) \tag{E13} \label{E13}
\end{equation}

Simplify:

\begin{equation}
    c_2^2 + c_3^2 = \frac{7}{8} - \frac{1}{2} c_1^2 - \frac{3}{2} c_4^2 \tag{E14} \label{E14}
\end{equation}

\subsection*{Step 9: Express \( (c_2 + 2 c_3)^2 \) in Terms of Known Quantities}

First, expand \( (c_2 + 2 c_3)^2 \):

\begin{align}
    (c_2 + 2 c_3)^2 &= c_2^2 + 4 c_2 c_3 + 4 c_3^2 \tag{E15} \label{E15}
\end{align}

From Eq.\eqref{E5}, rearranged:

\begin{equation}
    c_2 c_3 = -\sqrt{3}\, c_1 c_4 - \frac{3}{4} c_3^2 \tag{E16} \label{E16}
\end{equation}

Substitute \( c_2 c_3 \) into Eq.\eqref{E15}:

\begin{align*}
    (c_2 + 2 c_3)^2 &= c_2^2 + 4 \left( -\sqrt{3}\, c_1 c_4 - \frac{3}{4} c_3^2 \right ) + 4 c_3^2 \\
    &= c_2^2 - 4 \sqrt{3}\, c_1 c_4 - 3 c_3^2 + 4 c_3^2 \\
    &= c_2^2 - 4 \sqrt{3}\, c_1 c_4 + c_3^2 \tag{E17} \label{E17}
\end{align*}

Now, using \( c_2^2 + c_3^2 \) from Eq.\eqref{E14}:

\begin{equation}
  (c_2 + 2 c_3)^2  = \frac{7}{8} - \frac{1}{2} c_1^2 - \frac{3}{2} c_4^2 - 4 \sqrt{3}\, c_1 c_4\tag{E14}
\end{equation}

Thus, from Eq.\eqref{E12}, we have

\begin{align}
    &\left( 1 - c_1^2 - \left( \frac{7}{8} - \frac{1}{2} c_1^2 - \frac{3}{2} c_4^2 \right) - c_4^2 \right) c_4^2 \notag \\
    &= \frac{3}{7} c_1^2 \left( \frac{7}{8} - \frac{1}{2} c_1^2 - \frac{3}{2} c_4^2 - 4 \sqrt{3}\, c_1 c_4 \right) \tag{E18} \label{E18}
\end{align}

Simplifying we have

\begin{align*}
    & 28c_4^4 + \left(7 + 8c_1^2\right)c_4^2 + 96\sqrt{3}c_1^3c_4 + \left(12c_1^4 - 21c_1^2\right) = 0 \\
    \Leftrightarrow &\ \left(c_4 + \sqrt{3}c_1\right) \Big( 28c_4^3 - 28\sqrt{3}c_1 c_4^2 + \left(92c_1^2 + 7\right)c_4 \\
    &\qquad + \sqrt{3}\left(4c_1^3 - 7c_1\right) \Big) = 0
    \tag{E19} \label{E19}
\end{align*}

This then gives $c_4=-\sqrt{3}c_1$ as a solution.

\section{All ((7,2,3)) Stabilizer Codes}\label{app:all723stab}

The codeword stabilized (CWS) formalism provides a systematic approach for enumerating all \((7,2,3)\) stabilizer codes \cite{cross2008codeword}. A CWS code is characterized by a graph and a classical binary code. We examined both connected and disconnected graphs corresponding to \((7,2,3)\) stabilizer codes, taking into account graph isomorphisms and local unitary equivalence, as classified in \cite{hein2006entanglement}. This analysis results in 59 inequivalent graphs.

Assuming 0000000 is one of the classical codewords, we evaluated all 7-bit classical binary strings from 0000001 to 1111111 to identify the second codeword, selecting those that result in a CWS code with a minimum distance of 3. The resulting \((7,2,3)\) stabilizer codes are summarized in Table \ref{table:723-stab-code}, which lists the corresponding lengths of the signature vectors \(\lambda^*\).

\begin{table}[h!]
\centering
\begin{tabular}{c | c}
     \toprule
     $\lambda^*$ & Graph No. \\
     \midrule
     0 & 40, 42, 43, 44 \\
     $\sqrt{1}$ & 38, 39, 41 \\
     $\sqrt{2}$ & 17, 30, 35, 37 \\
     $\sqrt{3}$ & 8, (1, 8), 28, 33 \\
     $\sqrt{5}$ & (2, 4), 25, 31 \\
     \bottomrule
    \end{tabular}
\caption{\label{table:723-stab-code} Summary of all ((7,2,3)) stabilizer codes identified through the CWS formalism. Graph numbering follows \cite{hein2006entanglement}. For graphs with fewer than 7 vertices, isolated vertices were added, e.g., Graph No. 17. Notation (1, 8) refers to a disconnected graph composed of subgraphs No. 1 and No. 8.}
\end{table}

The signature vector for a stabilizer code can only contain components of 0 or 1, which implies that the square of its norm, \((\lambda^*)^2\), must be an integer. An exhaustive search reveals that for \((7,2,3)\) stabilizer codes, the only possible values of \(\lambda^*\) are \(\{0, \sqrt{1}, \sqrt{2}, \sqrt{3}, \sqrt{5}\}\).

\end{document}